\documentclass{IEEEtran}
\usepackage{cite}
\usepackage{amsmath,amssymb,amsfonts}
\usepackage{graphicx}
\usepackage{textcomp,nicefrac}
\usepackage{subcaption}
\usepackage{graphics}
\usepackage{epstopdf}
\usepackage{threeparttable} 
\usepackage[colorlinks=true,linkcolor=blue,citecolor=blue,urlcolor=blue]{hyperref}
\def\BibTeX{{\rm B\kern-.05em{\sc i\kern-.025em b}\kern-.08em
T\kern-.1667em\lower.7ex\hbox{E}\kern-.125emX}}
\begin{document}
	\title{Introducing Timepix2-Lite: A Miniaturized Readout Interface Enabling Nanosecond-Scale Half-Life Measurement}
\author{O. Pavlas, \IEEEmembership{Member, IEEE}, B. Bergmann, \IEEEmembership{Member, IEEE}, M. Holik, M. Malich, S. Pospisil, \IEEEmembership{Life Senior Member, IEEE}, P. Smolyanskiy, V. Vicha, R. Filgas
\thanks{This work has been funded by a grant from Programme Johannes Amos Comenius under the Ministry of Education, Youth and Sport of the Czech Republic [CZ.02.01.01/00/22\_008/0004590]. This work was supported in part within the frame of Horizon-Marie Skłodowska-Curie Actions (MSCA) Project Innovative Photodetector Module for Advanced Hybrid “Magnetic Resonance Imaging/Positron Emission Tomography” Scanners for Nuclear Medicine (INNMEDSCAN) as a part of Activities Related to Development of Front-End Electronics for Photonics Detectors under Agreement 101086178, and in part by the University of West Bohemia under Project SGS-2024-005. (Corresponding author: O. Pavlas.)}
\thanks{O. Pavlas is with Institute of Experimental and Applied Physics Czech Technical University in Prague (IEAP), 110 00 Prague, Czech Republic and also with Faculty of Electrical Engineering, Czech Technical University in Prague (CTU), 166 27 Prague, Czech Republic (e-mail: ondrej.pavlas@cvut.cz}
\thanks{B. Bergmann, M. Malich, S. Pospisil, P. Smolyanskiy, V. Vicha and R. Filgas are with IEAP, CTU
	in Prague, 110 00 Prague, Czech Republic (e-mail: benedikt.bergmann@cvut.cz, milan.malich@cvut.cz; stanislav.pospisil@cvut.cz; petr.smolyanskiy@cvut.cz; vladimir.vicha@cvut.cz; robert.filgas@cvut.cz).}
\thanks{M. Holik is with Faculty of Electrical Engineering, University of West Bohemia in Pilsen, 301 00 Pilsen, Czech Republic,
	and also with IEAP, CTU in Prague, 110 00 Prague, Czech Republic (e-mail:
	michael.holik.cz@gmail.com)
    }
    \thanks{The authors B.B. and P.S. acknowledge funding from the Czech Science Foundation under Registration Number GM23-04869M.}
}

\maketitle


\begin{abstract}
This paper presents Timepix2 Lite, a compact readout interface for the Timepix2 hybrid pixel detector, designed to provide high spatial and temporal resolution in a wide range of radiation detection applications. The Timepix2 Lite enables simultaneous energy and precise timing measurements on the nanosecond scale and supports advanced real-time data acquisition and visualization through the TrackLab software framework. Its compact and flexible architecture, enabled by the miniaturized design, allows seamless integration into diverse experimental setups, including nuclear spectroscopy, stratospheric balloon missions, and nanosecond-scale half-life measurements. As a case study, the half-life of the second excited state in $\mathbf{^{237}\mathbf{Np}}$ was determined to be $67.5(7)\,\mathrm{ns}$ for the 59.5~keV transition, consistent with previously reported results.

\end{abstract}

\begin{IEEEkeywords}
Timepix2, hybrid pixel detector, radiation detection, SESTRA, TrackLab, miniaturized readout, $^{241}$Am half life, $^{237}$Np.
\end{IEEEkeywords}

\section{Introduction}
\label{sec:introduction}
\IEEEPARstart {M}{odern} particle detection and radiation measurement require compact, flexible, and high-performance readout systems capable of precise temporal and spatial resolution. Hybrid pixel detectors, such as Timepix2~\cite{WONG2020106230}, support simultaneous Time-of-Arrival (ToA) and Time-over-Threshold (ToT) measurements, enabling characterization of fast radiation events.

\par The Timepix2 Lite interface builds upon the functionality of the Timepix2 chip, providing a miniaturized, USB-compatible platform that integrates seamlessly with the TrackLab~\cite{Mánek_2024}, \cite{TrackLab_software} software framework for advanced data acquisition and real-time visualization. 

\par To illustrate its capabilities, the Timepix2 Lite system was functionally tested at the Proton Synchrotron beam at CERN. In a separate experiment, it was employed to measure the half-life of the second nuclear excited state in $^{237}$Np using a time delayed-coincidence method. These experiments demonstrate that the system is capable of performing with accuracy of nanosecond-scale and also confirm its applicability in the field of nuclear physics, highlighting its potential for both laboratory-based research and field-deployable nuclear instrumentation.

\par This paper is organized as follows. Section~\ref{sec:Timepix2_Lite_readout} provides a detailed overview of the Timepix2 Lite readout, including its design features and functional testing. Section~\ref{sec:Measurment_half_life} presents the experimental validation of the system through nanosecond-scale half-life measurements of \textsuperscript{237}Np, and Section~\ref{sec:Discussion} discusses potential applications, performance considerations, and possible improvements for future deployments.

\section{Timepix2 Lite readout}
\label{sec:Timepix2_Lite_readout}
\begin{figure}[t!]
	\centering
	\includegraphics[width=3.5in]{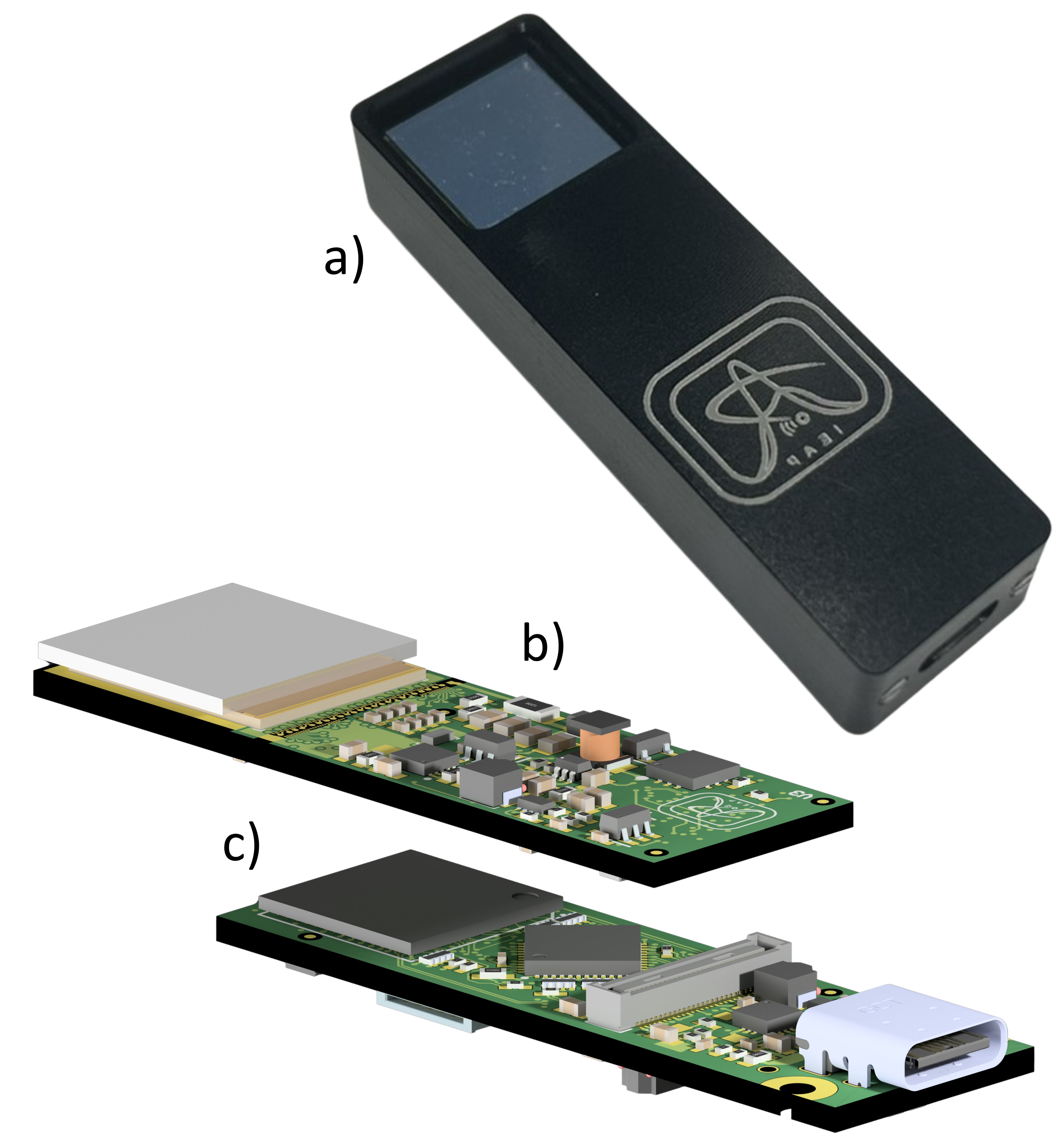}
   \caption{Timepix2 Lite readout: (a) overall view with dimensions 73 × 22 × 14 mm; the system consists of two printed circuit boards (PCBs): (b) the chipboard with the Timepix2 detector; and (c) the mainboard.}
	\label{Timepix2_Lite_real}
\end{figure}

The Timepix2 Lite is a new \cite{Pavlas2023}, miniaturized, highly compact and independently developed readout system specifically designed for radiation measurements. The Timepix2 Lite in its black anodized aluminum housing is shown in Fig.~\ref{Timepix2_Lite_real}(a). It builds on and fully exploits the capabilities of the hybrid pixel detector Timepix2, while offering substantial advantages in terms of size, usability, and system integration. The Timepix2 Lite enables straightforward connectivity through a single USB-C cable, simplifying experimental setup and significantly reducing cabling requirements in laboratory or field environments. Its compact dimensions (73~mm × 22~mm × 14~mm) and lightweight design (32~g) makes it particularly well-suited for applications where space or weight is limited, including mobile setups, educational demonstrations, or spaceborne instrumentation. Furthermore, the system provides a versatile platform capable of measuring both energy deposition and timing information simultaneously, which opens new possibilities for advanced experimental tasks. A detailed description of the capability for simultaneous measurement is provided in the following section.

\subsection{Timepix2}
\label{Timepix2}
Timepix2 is the successor to the Timepix \cite{LLOPART2007485} hybrid pixel detector, developed within the framework of the Medipix2 collaboration. The Timepix2 chip consists of a 256~×~256 pixel matrix with a pitch of 55~$\mu$m. Compared with the earlier Timepix detector, Timepix2 allows each activated pixel to measure the Time-over-Threshold (ToT) and the Time-of-Arrival (ToA) at the same time. 
\par The analog front-end of Timepix2 supports operation in adaptive gain mode, allowing per-pixel amplification adjustment. In this mode, a metal–oxide–semiconductor (MOS) capacitor is used in parallel with the feedback capacitor of the charge-sensitive amplifier (CSA) \cite{7286866}. Each pixel contains a 5-bit discriminator that allows fine adjustment of the threshold for each pixel across the entire chip. The analog front-end parameters are configured globally through 19 internal programmable DACs (Digital to Analog Converts). These individual DACs can be measured via the DACOUT (DAC Output) pin of the Timepix2. The DAC to be measured is selected through the internal DACOUTSEL (DAC Output Select) register. In addition, each pixel can be individually masked at the analog level; in this mode, the current source to the discriminator is disabled and the current supplied to the CSA is reduced from several microamperes to a few nanoamperes \cite{WONG2020106230}. This feature is critical for low-power applications, such as space instrumentation, radiation dosimetry, etc.
\par The digital front-end of Timepix2 provides adaptability across multiple applications. Each pixel contains four counters: two 10-bit counters (A and B) and two 4-bit counters (C and D), which can be combined to form extended counters depending on the operating mode. Timepix2 supports both simultaneous and continuous acquisition modes. In simultaneous mode, energy (ToT) and timing (ToA) information are recorded in parallel, with sequential readout determining the effective dead time, measured at approximately 18.5~ms with a 100~MHz data clock. In continuous read/write (CRW) mode, the counters alternate between acquisition and readout. This mechanism minimizes dead time, which is determined by the time required to read one set of counters (10 or 14~bits) and ranges from 6.6~ms to 9.2~ms at a data clock speed of 100~MHz. The clocking scheme further enhances flexibility. While the ToT clock is supplied externally by the readout system, the ToA signal is generated internally from the ToT base clock using a programmable divider. This allows independent optimization of energy and timing performance, e.g., employing a fast ToT clock for high-resolution energy measurements in combination with a slower ToA clock to extend the acquisition window. To support autonomous operation in rapidly changing radiation environments, a Matrix Occupation Monitor is implemented, triggering when the occupancy of a predefined pixel column region of Timepix2 exceeds a set threshold. It means the readout system can terminate the acquisition process once a predefined number of columns have been activated during the current acquisition. This prevents overexposure of the frame and overlap of tracks.

\begin{figure}[t]
	\centering
	\includegraphics[width=3.5in]{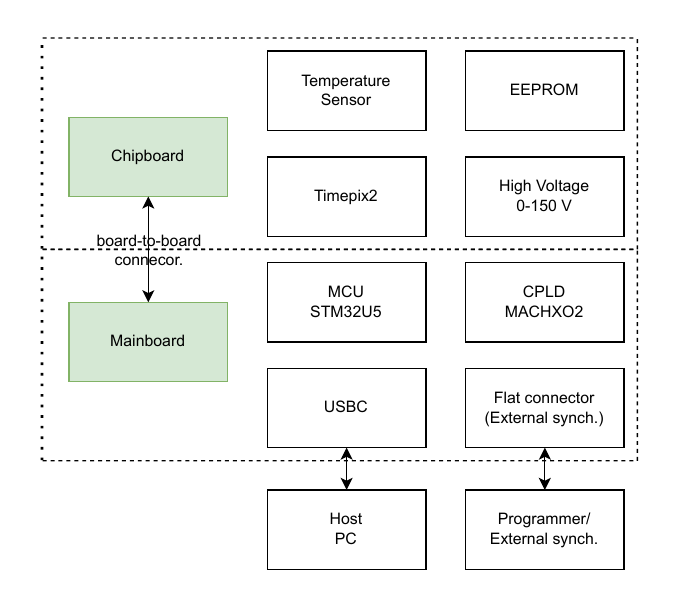}
	\caption{Readout hardware descriptor of Timepix2 Lite.}
	\label{HW_descriptor}
\end{figure}

\subsection{Readout system description}
The development of Timepix2 Lite represents the next step in the evolution of readout systems designed for hybrid pixel detectors. Building on the functionality of earlier devices, Timepix2 Lite has been introduced to provide a compact, cost-effective solution while maintaining compatibility with the Timepix2 architecture. It is the first miniaturized readout interface based on Timepix2 integrated into the TrackLab framework. Previous readout systems from the same family include the Katherine Readout Interface for Timepix2/3 \cite{Burian_2020}, \cite{Burian_2017}, Hardpix \cite{FILGAS2022620}, USB Lite \cite{VYKYDAL2011S48}, FITPix \cite{VKraus_2011} and MiniPIX SPRINTER \cite{advacam_minipix_sprinter}, each developed to address specific experimental needs ranging from laboratory characterization to portable field applications.

\par
\begin{figure}[t]
	\centering
	\includegraphics[width=3.5in]{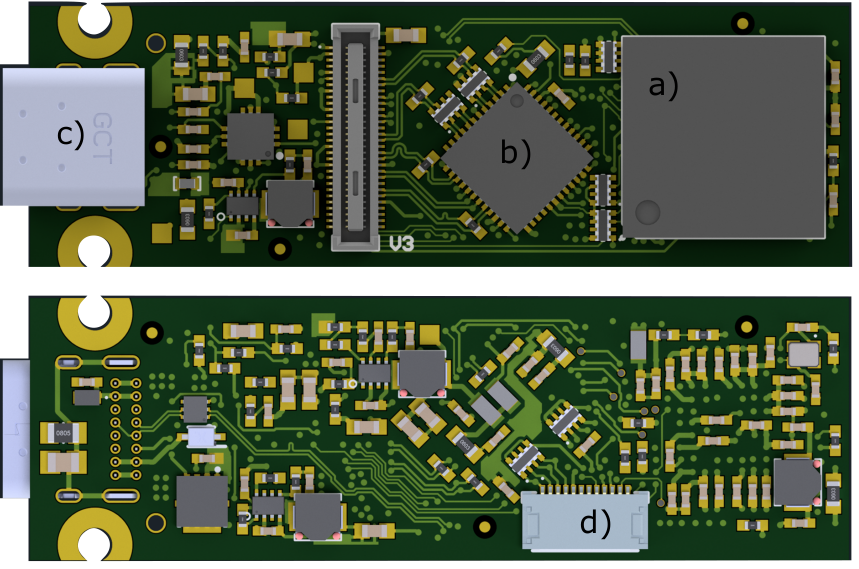}
	\caption{Detailed top and bottom views of the mainboard PCB. Labels (a-d) indicate the main components.} 
	\label{fig:mainboard}
\end{figure}

The core of the mainboard, shown as printed circuit board~(c) in Fig.~\ref{Timepix2_Lite_real}, is built around a microcontroller (MCU) STM32U5A9, which serves as the primary control unit of the readout system. Its main role is to manage communication between the host PC and the Timepix2 detector. In addition, a Complex Programmable Logic Device (CPLD) is integrated on the mainboard to perform logic-level translation between 3.3~V single-ended CMOS and differential SLVS standards, and vice versa. An external 32~MHz oscillator is used as the primary clock input to the MCU, from which the output clock is derived via a Phase-Locked Loop (PLL) and subsequently provided to the Timepix2 detector for measurements.
For monitoring the DACOUT signal of the Timepix2 detector, the internal 12-bit Analog-to-Digital Converter (ADC) of the MCU is used. Data readout from the Timepix2 is performed via a serial communication interface with a maximum clock frequency of 80~MHz. In the ToT10/ToA18 operating mode, reading out the complete pixel matrix requires 23~ms. The acquired data are subsequently processed by the on-board MCU and transmitted to the host computer; this step was measured to take approximately 10~ms during measurements with an Am$^{241}$ source. Consequently, the maximum achievable frame rate is limited to 30 frames per second. External synchronization and system programming are provided through a 12~pin flat FFC connector with pitch 0.5~mm located on the side of the device, allowing easy integration with additional hardware or experimental setups. This connector is available after opening the aluminum housing for applications requiring external synchronization. The mainboard and chipboard are interconnected via a board-to-board connector featuring a 0.4~mm pitch and a 2.5~mm stacking height. Mainboard components described in the previous part are illustrated in Fig.~\ref{fig:mainboard}, the labels (a-d) summarize the main components described above:
a) MCU STM32U5A9, 
b) CPLD MachXO2, 
c) USBC, 
and d) Flat connector.

\par
\begin{figure}[t]
	\centering
	\includegraphics[width=3.5in]{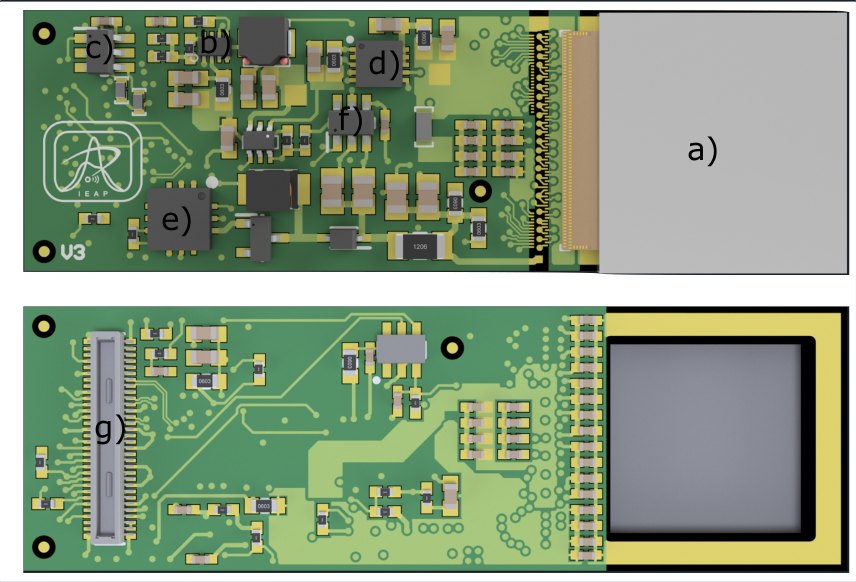}
	\caption{Detailed top and bottom views of the chipboard PCB. Labels (a-f) indicate the main components.} 
	\label{fig:chipboard}
\end{figure}

 
The chipboard, shown as the printed circuit board (b) in Fig.~\ref{Timepix2_Lite_real}, houses the Timepix2 detector equipped with a 500~µm silicon sensor layer, forming the core of the radiation detection system. A positive high voltage, adjustable from 0 to 150~V, is supplied to the detector through an onboard integrated high-voltage source, which is based on MAX1932 component, providing biasing for semiconductor sensor operation. In addition, an EEPROM (Electrically Erasable Programmable Read-only Memory) is mounted on the chipboard to store key chipboard information, including the detector type, operational properties, and the Media Access Control (MAC) address of the entire system, facilitating identification and configuration in experimental setups.
A 12-bit digital temperature sensor, TMP100, is positioned in close proximity to the Timepix2 detector, enabling local temperature measurements in the range of $-55$ to $125^\circ$C with a precision of $\pm 1^\circ$C. These measurements can be correlated with the internal temperature readings provided by the Timepix2 itself, allowing for thermal monitoring and compensation during operation. Separate power domains for the analog and digital sections of the detector are implemented on the chipboard. The analog supply is provided by an ultra-low-noise low-dropout voltage regulator TPS7A94, while the digital supply is generated by a step-down converter MP2333H. And Chipboard components described in the previous part are illustrated in Fig.~\ref{fig:chipboard}, the labels (a-f) summarize the main components described above:
a) Timepix2, 
b) Step-down convert MP2333H, 
c) EEPROM, 
d) LDO TPS7A94, 
e) HV MAX1932, 
f) Temperature sensor TMP100, 
and g) Board-to-board connector.

\par
Both the chipboard and the mainboard are housed in a mechanical aluminum enclosure with a black anodized finish, featuring openings for the Timepix2 sensor and the USBC user interface connector.

\begin{table}
\centering
\caption{\small\textsc{Parameters of the readout system Timepix2 Lite}}
\label{tab:tpx2lite}
\renewcommand{\arraystretch}{1.3} 
\begin{threeparttable}
\begin{tabular}{l l}
\hline
\textbf{Timepix2 Lite} &  \\
\hline
Dimensions:        & (73 $\times$ 22 $\times$ 14) mm \\
Weight:             & 32 g \\
Interface:        & USB 2.0 Type C \\
Timepix2 Readout Speed:    & 80 MHz \\
Power consumption\tnote{1}: & 1.5~W \\
High Voltage:     & 0 -- 150~V \\
ToA Resolution:   & 12.5 ns \\
Frame Rates\tnote{2}: & 30~fps (Mode: ToT10/ToA18)\\
Software:          & TrackLab \cite{Mánek_2024}, \cite{TrackLab_software}\\
Tested Temperature Ranges: & -20$^{\circ}$~C up to 70$^{\circ}$~C \\
Architecture Based: & MCU \\
\hline
\end{tabular}
\begin{tablenotes}
\item[1] Measured during measurement with Am$^{241}$. Timepix2 mode: ToT10/ToA18 and 80~MHz data clock.
\item[2] Measured using 80~MHz data clock.
\end{tablenotes}
\end{threeparttable}
\end{table}

\subsection{Communication with computer and SW integration}
Timepix2 Lite, when connected to a host computer via USB-C, is enumerated as a USB Remote Network Driver Interface Specification (RNDIS) class device, enabling seamless plug-and-play connectivity. Data transfer between the device and the host is carried out using the User Datagram Protocol (UDP) over USB High-Speed (2.0), in full compliance with the standards specified in \cite{Burian_2017}.
At the hardware level, the system performs extensive preprocessing of the pixel data acquired from the Timepix2 detector. This processing includes deserialization and derandomization operations. Deserialization is required because the Timepix2 does not transmit data on a pixel-by-pixel basis, but rather in a dedicated serial format. Derandomization is necessary because the Timepix2 utilizes linear feedback shift register (LFSR) counters for time and energy measurements. The derandomization is implemented using a look-up table. The entire device is fully integrated into the in-house software framework TrackLab, which provides real-time data processing, visualization, analysis, and clustering functionalities, among other capabilities. By leveraging TrackLab, users gain full access to the advanced features of the Timepix2 detector, including live monitoring of pixel hits, energy deposition, and timing information. This integration simplifies experimental workflows and enables both laboratory-based research and field-deployable nuclear instrumentation to fully exploit the detector’s capabilities. Technical parameters of the Timepix2 Lite are provided in Table \ref{tab:tpx2lite}.

\subsection{Functional Testing of the Timepix2 Lite at the CERN Super Proton Synchrotron}
\begin{figure}[t]
	\centering
	\includegraphics[width=2.5in]{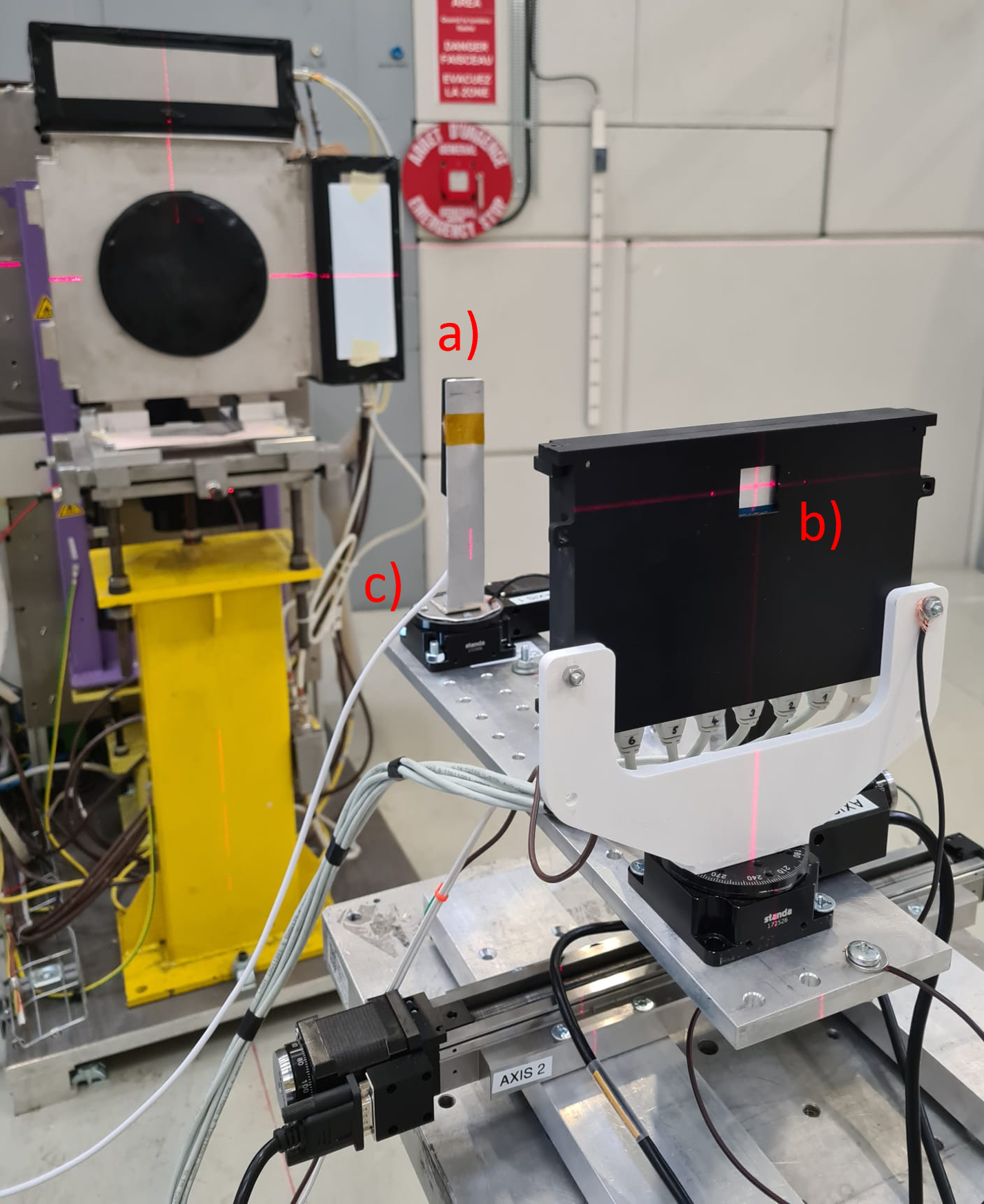}
	\caption{Experimental setup of the Timepix2 Lite at the CERN Super Proton Synchrotron. The backside of the Timepix2 Lite (a) is mounted on a rotation stage (c), and the center of the incoming hadron beam is indicated by (b).}
	\label{fig:setup}
\end{figure}

Timepix2 Lite was tested at the beamline of the Super Proton Synchrotron at CERN to evaluate its functional performance under realistic high-energy particle conditions. The entire setup was mounted on a precision rotation stage, allowing fine adjustment of the detector orientation relative to the incoming beam in steps of $1^\circ$. The entire setup is shown in Fig.~\ref{fig:setup}. This configuration enabled systematic measurements at multiple angles to study the angular dependence of detector response. The beam at Super Proton Synchrotron at CERN consisted of hadrons with an energy of 180~GeV/c. 

\par
\begin{figure}[t]
    \centering

    \begin{subfigure}[b]{0.45\columnwidth}
        \includegraphics[width=1.1\linewidth]{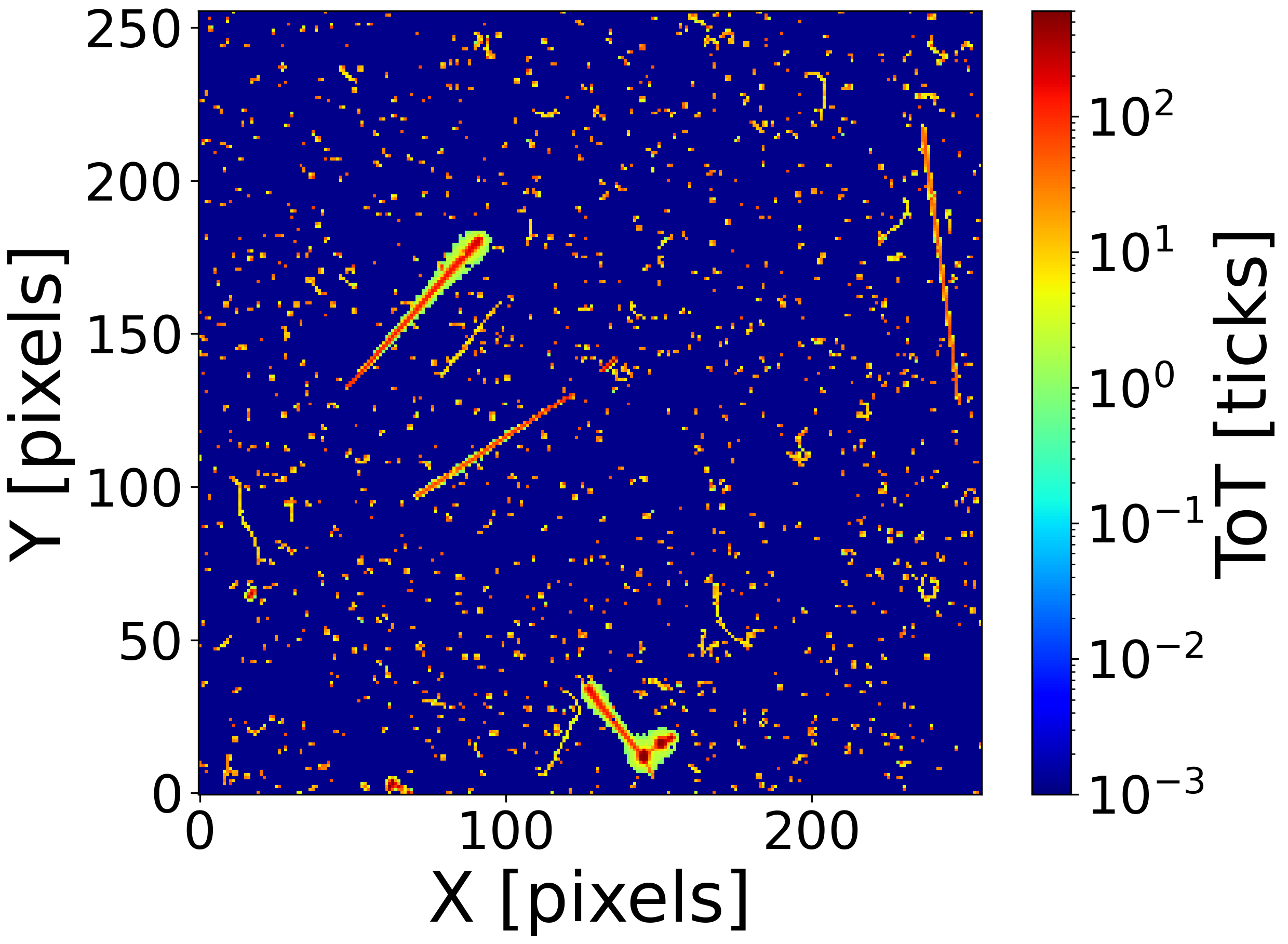}
    \end{subfigure}
    \hspace{0.022\columnwidth}
    \begin{subfigure}[b]{0.45\columnwidth}
        \includegraphics[width=1.1\linewidth]{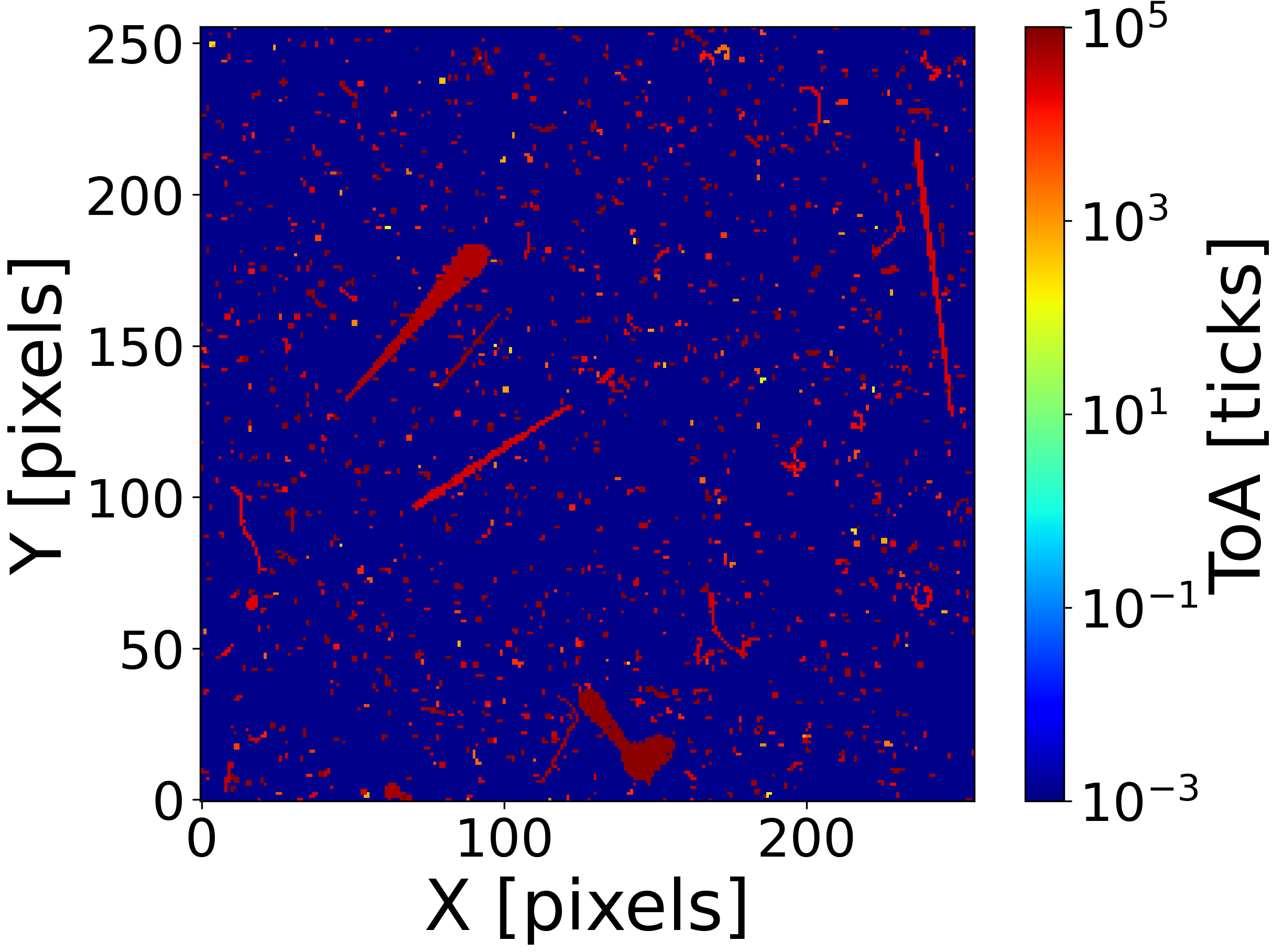}
    \end{subfigure}

    \vspace{0cm}
    \centering{\small a) Rotation 0$^{\circ}$}

    \vspace{0.35cm}

    \begin{subfigure}[b]{0.45\columnwidth}
        \includegraphics[width=1.1\linewidth]{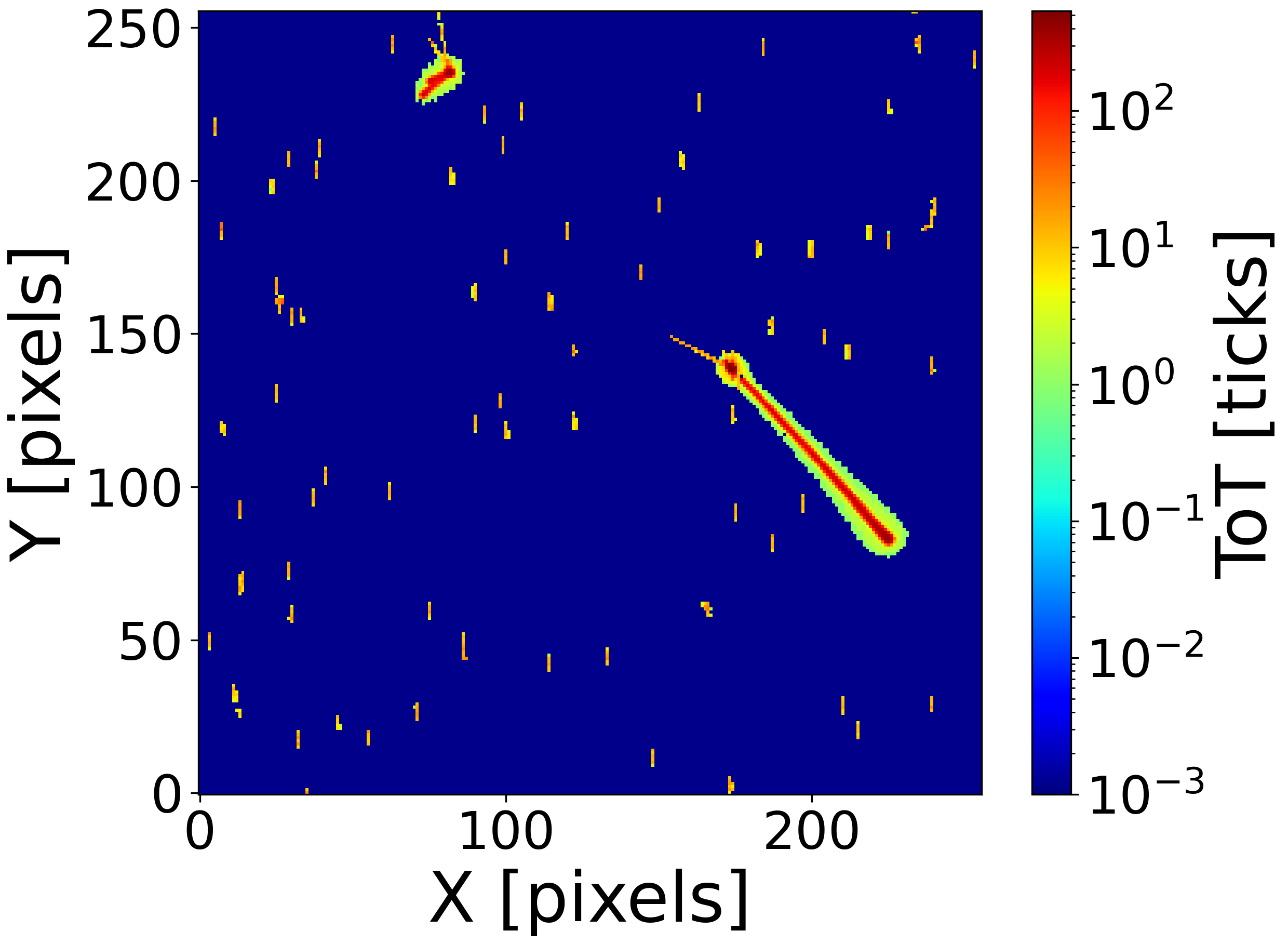}
    \end{subfigure}
    \hspace{0.022\columnwidth}
    \begin{subfigure}[b]{0.45\columnwidth}
        \includegraphics[width=1.1\linewidth]{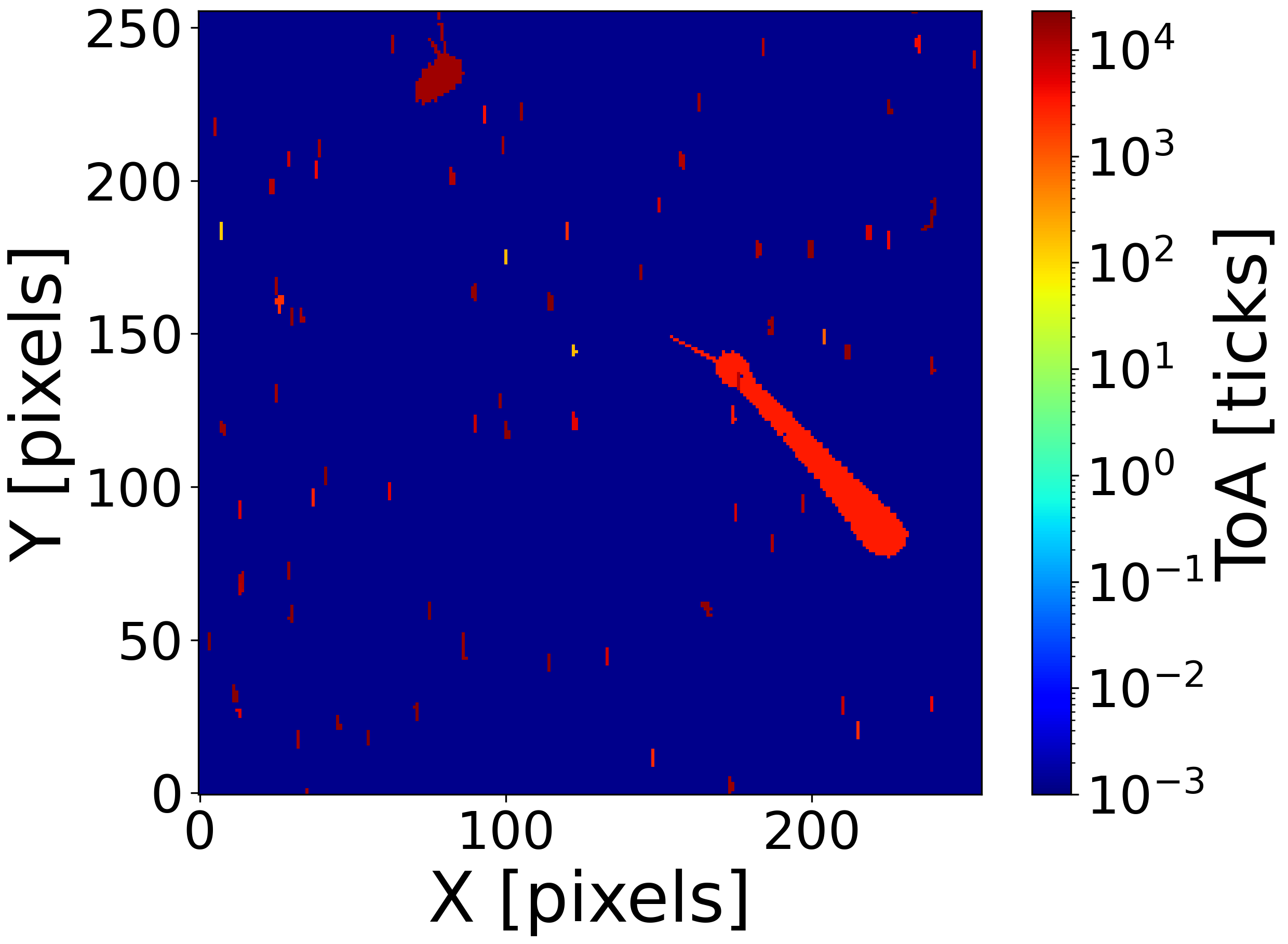}
    \end{subfigure}

    \vspace{0cm}
    \centering{\small b) Rotation 30$^{\circ}$}

    \vspace{0.35cm}

    \begin{subfigure}[b]{0.45\columnwidth}
        \includegraphics[width=1.1\linewidth]{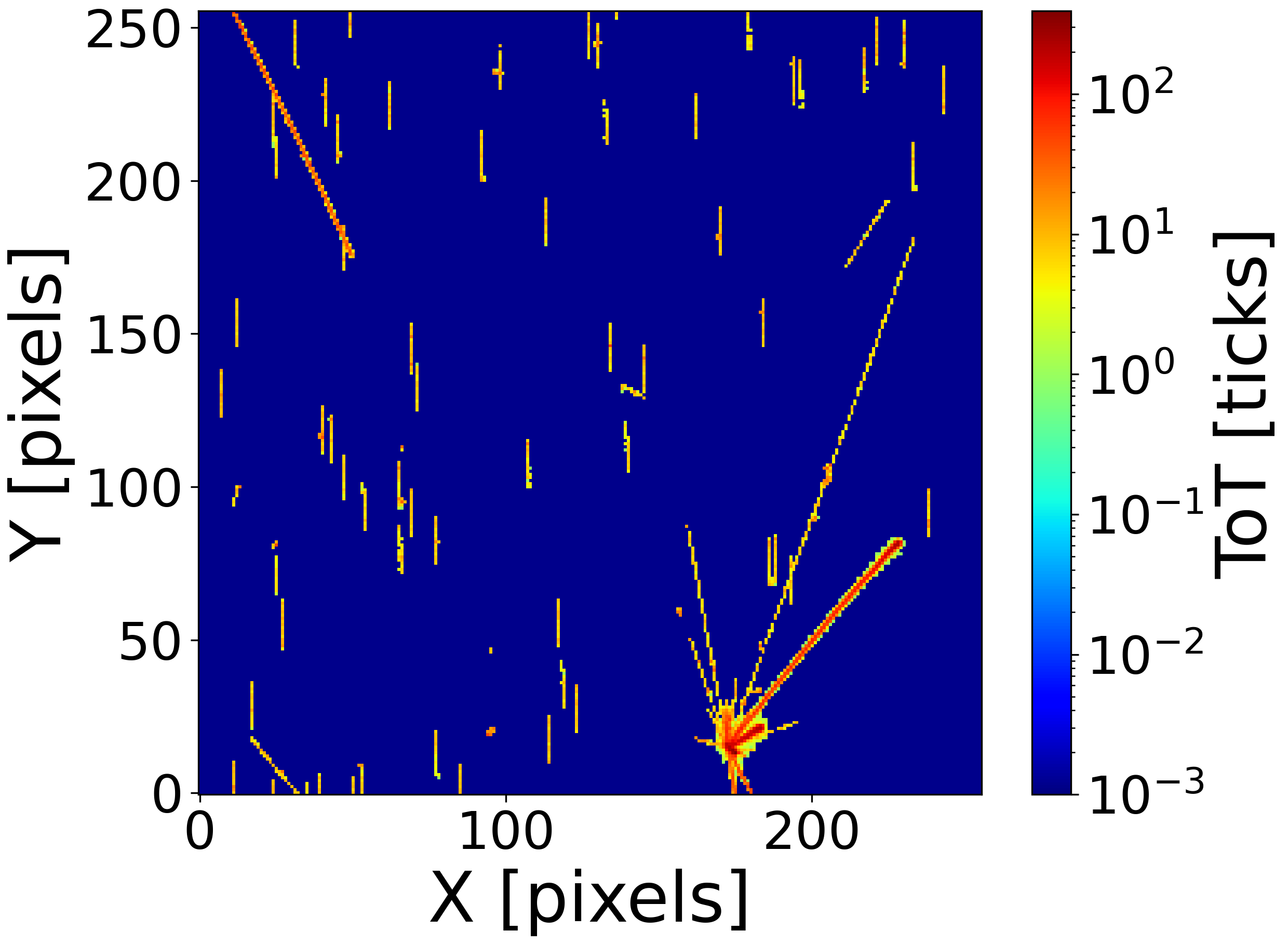}
    \end{subfigure}
    \hspace{0.022\columnwidth}
    \begin{subfigure}[b]{0.45\columnwidth}
        \includegraphics[width=1.1\linewidth]{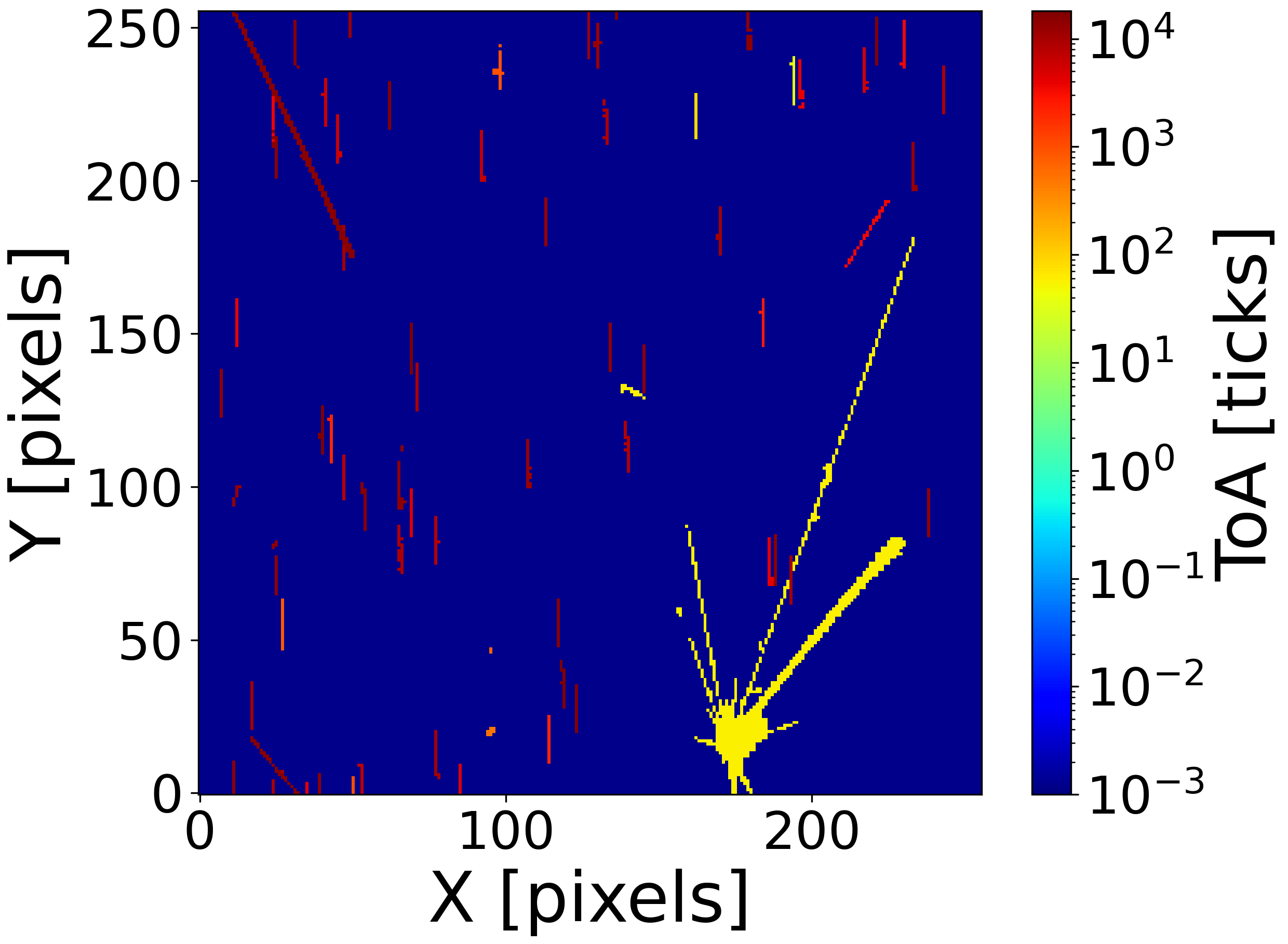}
    \end{subfigure}

    \vspace{0cm}
    \centering{\small c) Rotation 60$^{\circ}$}

    \vspace{0.35cm}

    \begin{subfigure}[b]{0.45\columnwidth}
        \includegraphics[width=1.1\linewidth]{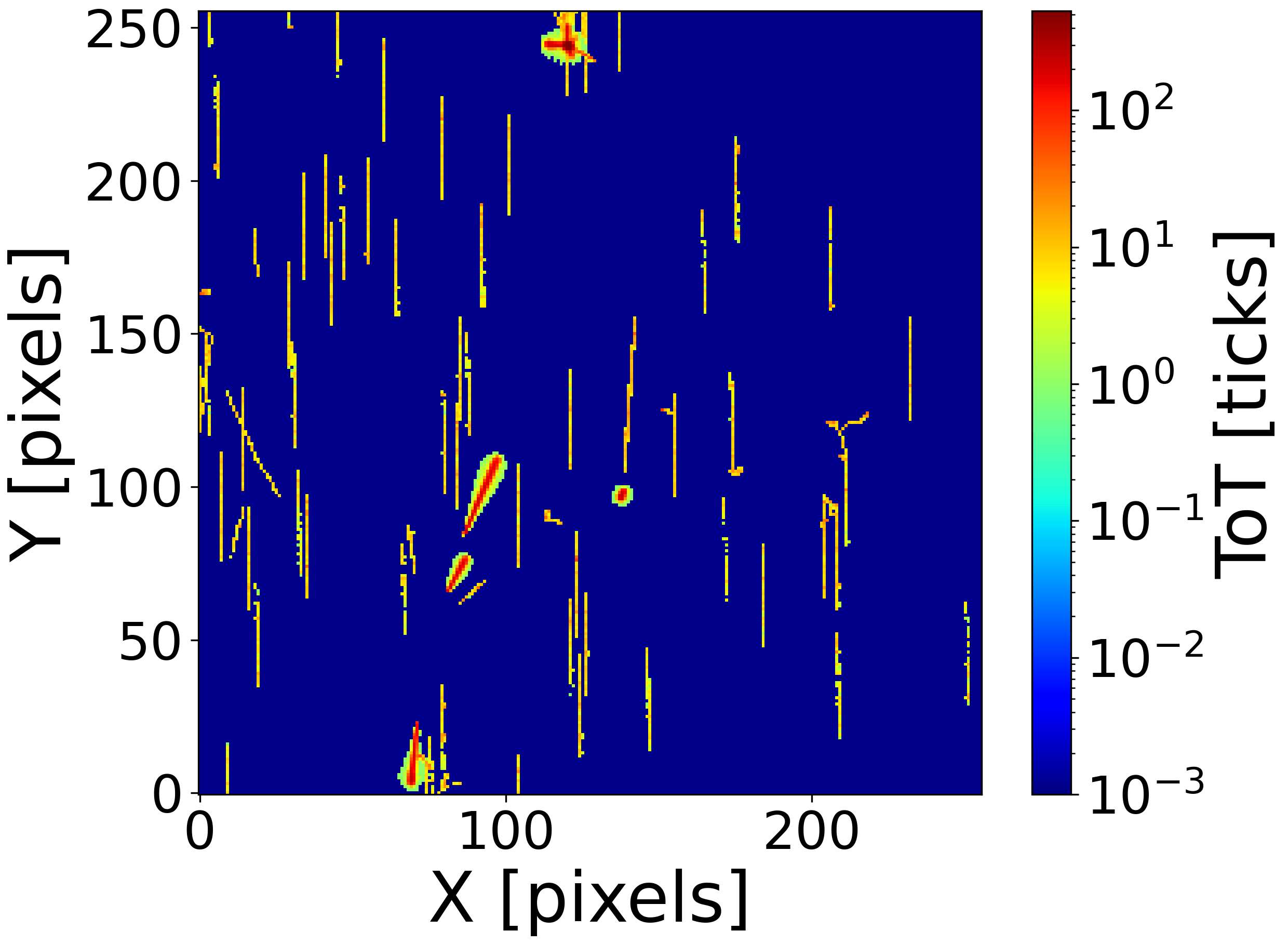}
    \end{subfigure}
    \hspace{0.022\columnwidth}
    \begin{subfigure}[b]{0.45\columnwidth}
        \includegraphics[width=1.1\linewidth]{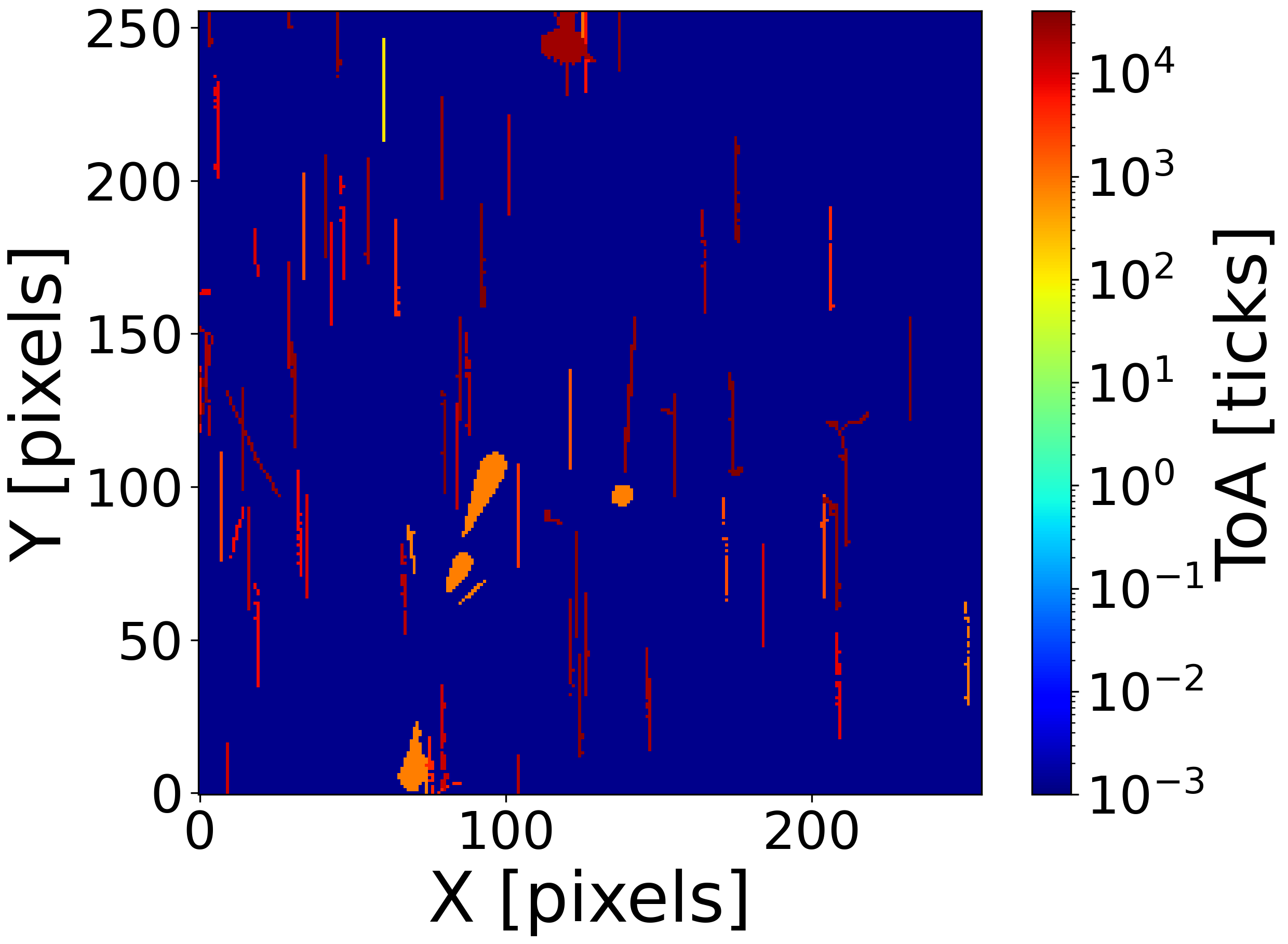}
    \end{subfigure}

    \vspace{0cm}
    \centering{\small d) Rotation 75$^{\circ}$}

    \vspace{0.4cm}
    \begin{minipage}[t]{0.45\columnwidth}
        \centering
        \small Time-over-Threshold (ToT)
    \end{minipage}
    \hspace{0.022\columnwidth}
    \begin{minipage}[t]{0.45\columnwidth}
        \centering
        \small Time-of-Arrival (ToA)
    \end{minipage}

    \caption{Time-over-Threshold (ToT, left) and Time-of-Arrival (ToA, right) responses of the Timepix2 detector exposed to hadrons at various incidence angles in the CERN Super Proton Synchrotron beam.}
    \label{fig:tot_toa_combined}
\end{figure}

 Representative frames acquired by the Timepix2 Lite at different orientations are shown in Fig.~\ref{fig:tot_toa_combined}, illustrating the detector’s capability to record particle tracks with high spatial resolution, defined by a pixel pitch of 55~$\mu$m. Timepix2 was operated in simultaneous ToT10/ToA18 mode, using both adaptive and normal gain settings. The left side of Fig.~\ref{fig:tot_toa_combined} shows frames displaying the ToT response from the measurement. Primary particles can be distinguished from secondaries using the Time-of-Arrival information. The right side of Fig.~\ref{fig:tot_toa_combined} shows the corresponding ToA frames, where primary and secondary particles are identified based on their different arrival times. These frames highlight the effect of varying the detector angle on track shape and energy deposition, and also demonstrate the ability to separate tracks not only based on their spatial properties but also using their arrival times.

\par
\begin{figure}[t]
    \centering
    \begin{subfigure}[b]{0.45\columnwidth}
        \makebox[\linewidth][c]{%
            \includegraphics[width=1.2\linewidth]{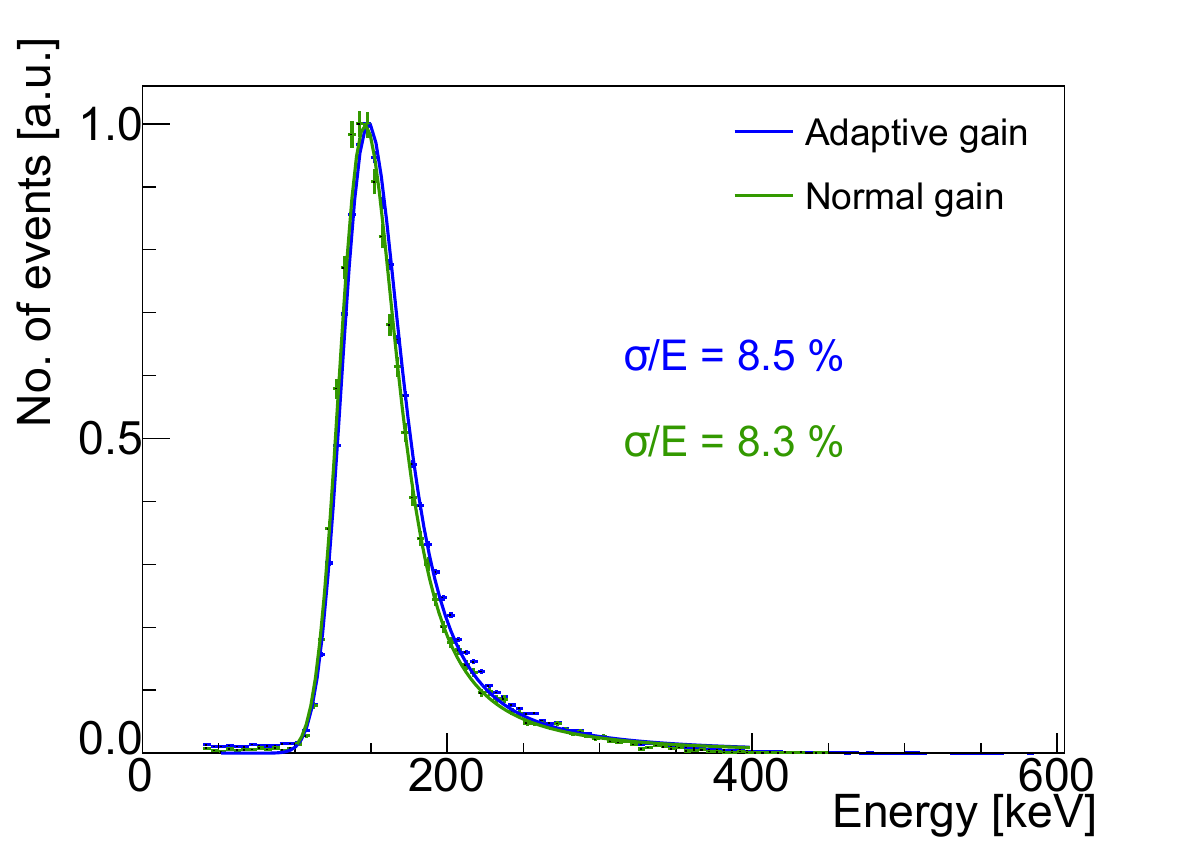}%
        }
        \caption{Rotation $0^{\circ}$}
        \label{fig:0a}
    \end{subfigure}
    \hspace{0.05\columnwidth}
    \begin{subfigure}[b]{0.45\columnwidth}
        \makebox[\linewidth][c]{%
            \includegraphics[width=1.2\linewidth]{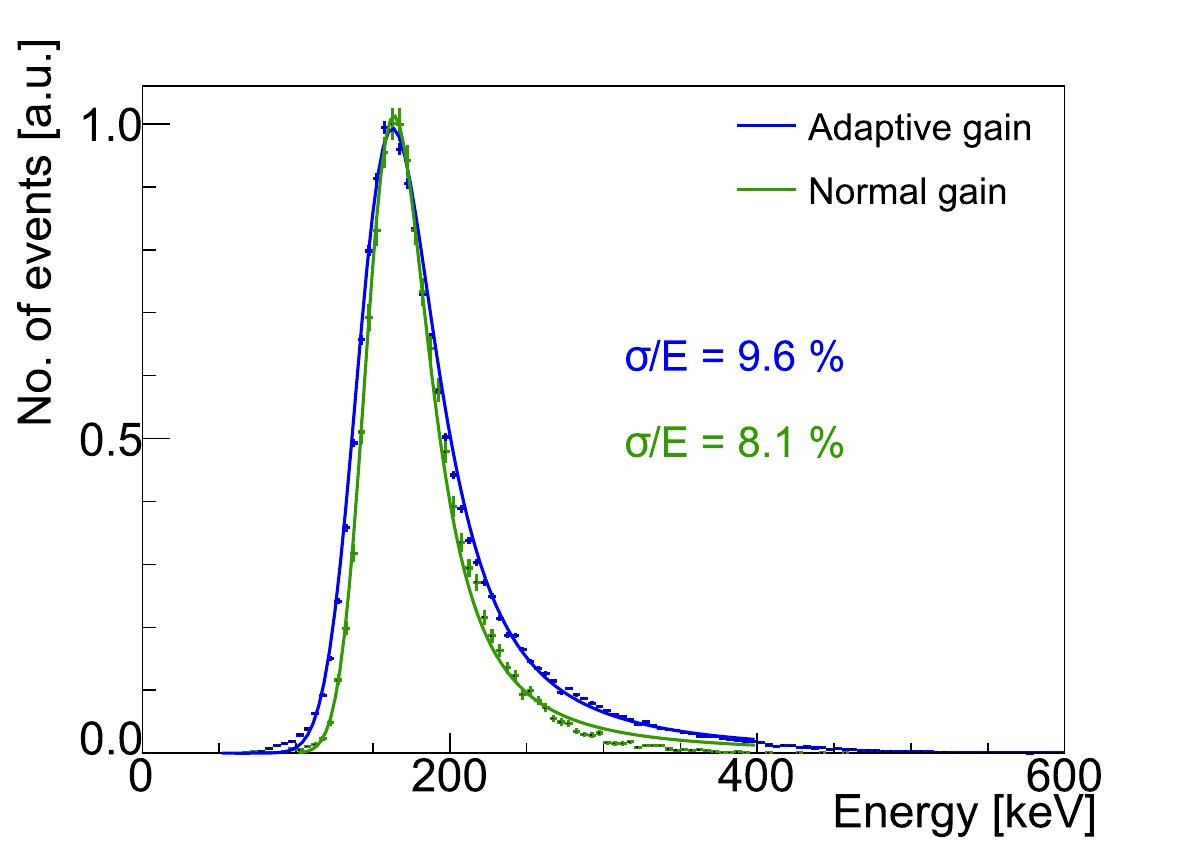}%
        }
        \caption{Rotation $30^{\circ}$}
        \label{fig:30a}
    \end{subfigure}

    \vspace{0.3cm}

    \begin{subfigure}[b]{0.45\columnwidth}
        \makebox[\linewidth][c]{%
            \includegraphics[width=1.2\linewidth]{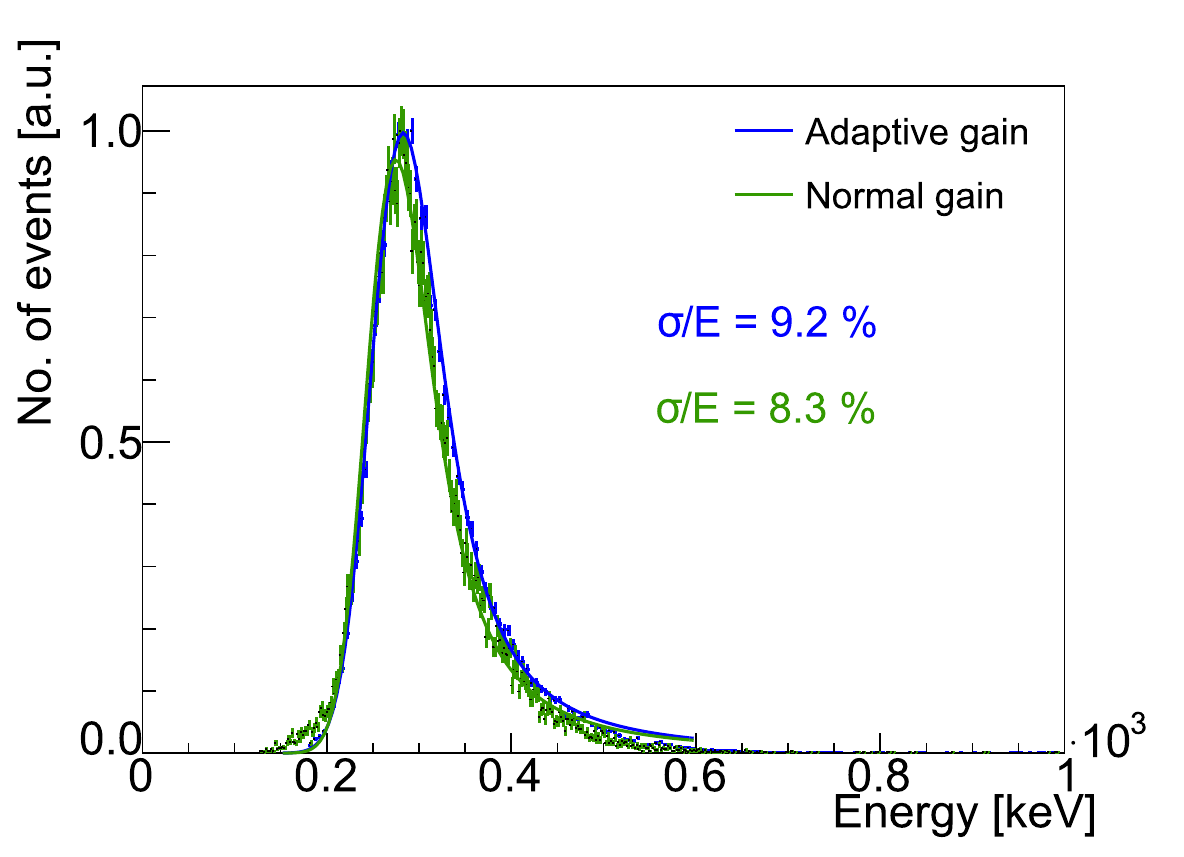}%
        }
        \caption{Rotation $60^{\circ}$}
        \label{fig:60a}
    \end{subfigure}
    \hspace{0.05\columnwidth}
    \begin{subfigure}[b]{0.45\columnwidth}
        \makebox[\linewidth][c]{%
            \includegraphics[width=1.2\linewidth]{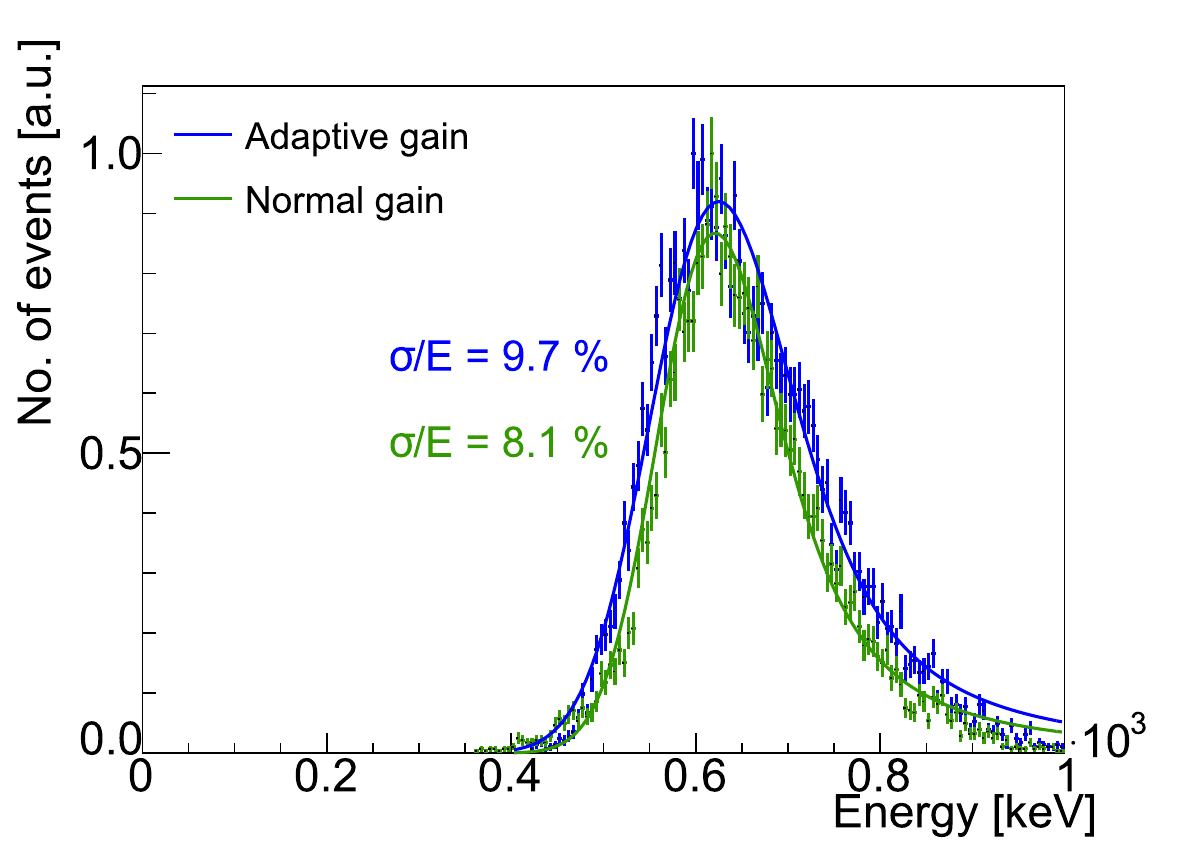}%
        }
        \caption{Rotation $75^{\circ}$}
        \label{fig:75a}
    \end{subfigure}

    \caption{Comparison of energy deposition spectra measured at different hadron impact angles relative to the sensor normal, obtained in both adaptive and normal gain modes. The spectra were fitted with a Landau convoluted with a Gaussian function.}
    \label{fig:tracksa}
\end{figure}

 The dependence of the deposited energy on the detector’s rotation angle is quantitatively presented in Fig.~\ref{fig:tracksa}. Measurements were performed in both the adaptive and normal gain modes of Timepix2 to evaluate the effect of gain settings on detector performance. At all angles, both modes provide the relative energy resolution ($\frac{\sigma}{E}$) of ~8\%. It also shows that, with increasing rotation angle relative to the sensor normal, the energy deposited in the sensor layer increases.

\section{Measurement of second nuclear excited state in $^{237}$Np}
\label{sec:Measurment_half_life}
The measurement of radioactive half-lives has traditionally been a demanding instrumentation task, typically requiring specialized laboratory settings \cite{Vretenar_2019, DUTSOV2021, SANTOS2023}. By employing the Timepix2 Lite detector within the new generation of SEST$^2$RA kit (School Education Set with Timepix2 for Radiation Analysis) (see Fig.~\ref{fig:Experimental_setup}), which succeed the original SESTRA kits (School Education Set with Timepix2 for Radiation Analysis) \cite{Vicha2017} and incorporates the capabilities of the Timepix2 Lite readout. It is now possible to perform radioactive half-life measurements on a compact tabletop setup. This approach lowers technical barriers while enabling precise investigation of radioactive decay processes, with a measurement precision on the nanosecond timescale for half-life determination. A representative application of the detector is the measurement of the half-life of the second nuclear excited state in $^{237}$Np, performed using the delayed-coincidence method, demonstrating its capability for nanosecond-scale timing measurements in nuclear physics experiments.

\subsection{Experimental setup}
\label{Experimental setup}
\begin{figure}[t]
	\centering
	\includegraphics[width=3.5in]{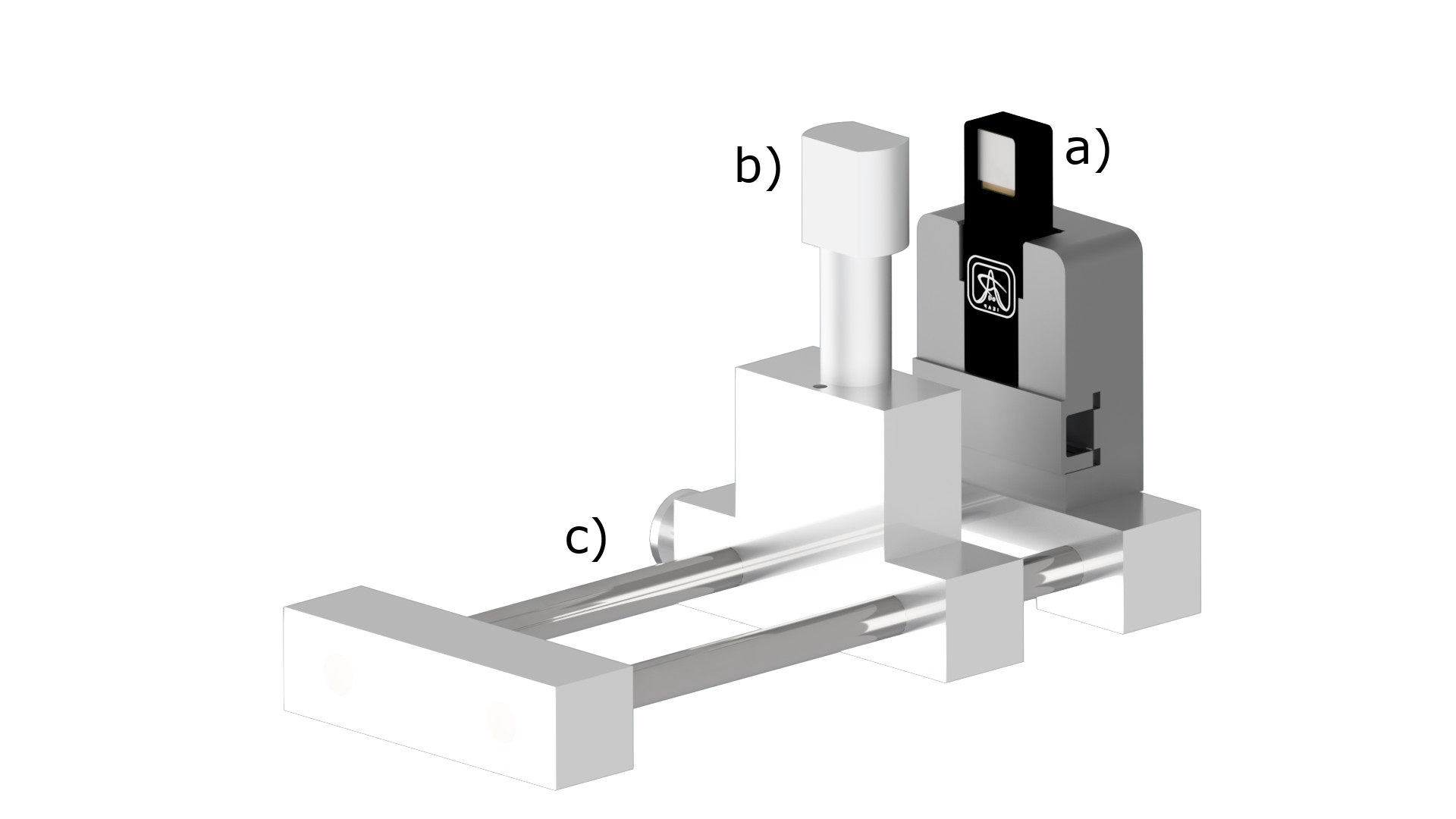}
	\caption{Experimental setup with SEST$^2$RA kit for measuring the half-life of the second excited state of $^{237}$Np in air using (a) a Timepix2 Lite detector, (b) $^{241}$Am source and (c) laboratory bench.} 
	\label{fig:Experimental_setup}
\end{figure}
The SEST$^2$RA kit was used to measure the half-life of the second excited state in $^{237}$Np. The experimental setup, including the laboratory bench, the Timepix2 Lite detector, and an Eckert \& Ziegler CESIO $^{241}$Am source with an activity of 9.5~kBq, is shown in Fig.~\ref{fig:Experimental_setup}. For the experiment, the source was operated with a wide output and positioned in air at a distance of 1.5~cm from the top surface of the silicon sensor layer of the Timepix2 Lite detector. 
\par The Timepix2 detector was configured in the digital ToT10/ToA18 mode, allocating 10~bits for Time-over-Threshold (ToT) and 18~bits for Time-of-Arrival (ToA), with sequential read/write. The ToT measurement, which provides information about the deposited energy in each pixel, was driven by an input clock with a frequency of 80~MHz, ensuring temporal resolution of 12.5~ns for the expected signals. No internal clock division was applied for the ToA measurement, allowing the full 18-bit counter to record event arrival times over a maximum range of 3.3~ms. The acquisition time for each frame was set to 3~ms, providing sufficient statistics for reliable energy and timing analysis. A high voltage of 150~V was applied to the sensor layer. The per-pixel energy calibration of the detector was performed prior to the measurement using X-ray fluorescence lines, following the procedure described in \cite{JAKUBEK2011S262}, and a single calibration was used for the entire experiment. For the coincidence measurement, the threshold was increased from 5~keV (the calibrated value) to 8~keV to minimize the number of noisy pixels in the measured data. This adjustment is acceptable because the lowest energy considered in the half-life coincidence analysis is 59.5~keV.

\subsection{Decay Scheme of $^{241}$Am}
\begin{figure}[t]
	\centering
	\includegraphics[width=3.5in]{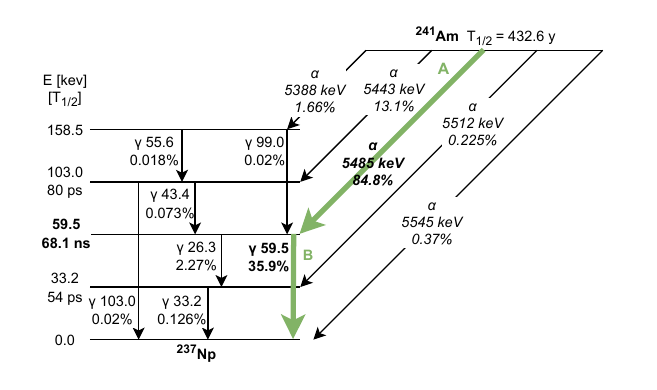}
	\caption{Decay scheme of $^{241}$Am, showing energy levels relevant for half-life measurements. Transition A corresponds to the $\alpha$-particle, and transition B corresponds to the $\gamma$-photon.}
	\label{Am241_decay}
\end{figure}

The decay scheme of $^{241}$Am \cite{Am241Table} is illustrated in Fig.~\ref{Am241_decay}. In this process, the $^{241}$Am nucleus undergoes $\alpha$-decay to form $^{237}$Np, emitting $\alpha$-particles with several discrete energies. Each $\alpha$-particle energy corresponds to a transition to a specific excited state of the daughter nucleus. The relative intensities of these transitions determine the probability of populating each state, which is important for understanding the energy spectrum measured by the detector. This decay scheme provides the foundation for experiments involving $\alpha$-spectroscopy and $\gamma$-spectroscopy for half-life measurements of the second excited states in $^{237}$Np. Population of the 59.5~keV level by unresolved alpha decays (5388~keV and 5443~keV), via the decay of the 158.5~keV and 103.0~keV levels, also includes measurements of the half-lives of these levels. As shown in Fig. \ref{Am241_decay}, the half-lives of these levels are on the order of picoseconds, and therefore this systematic error can be neglected.

\subsection{Data Analysis}
\begin{figure}[t]
	\centering
	\includegraphics[width=3.5in]{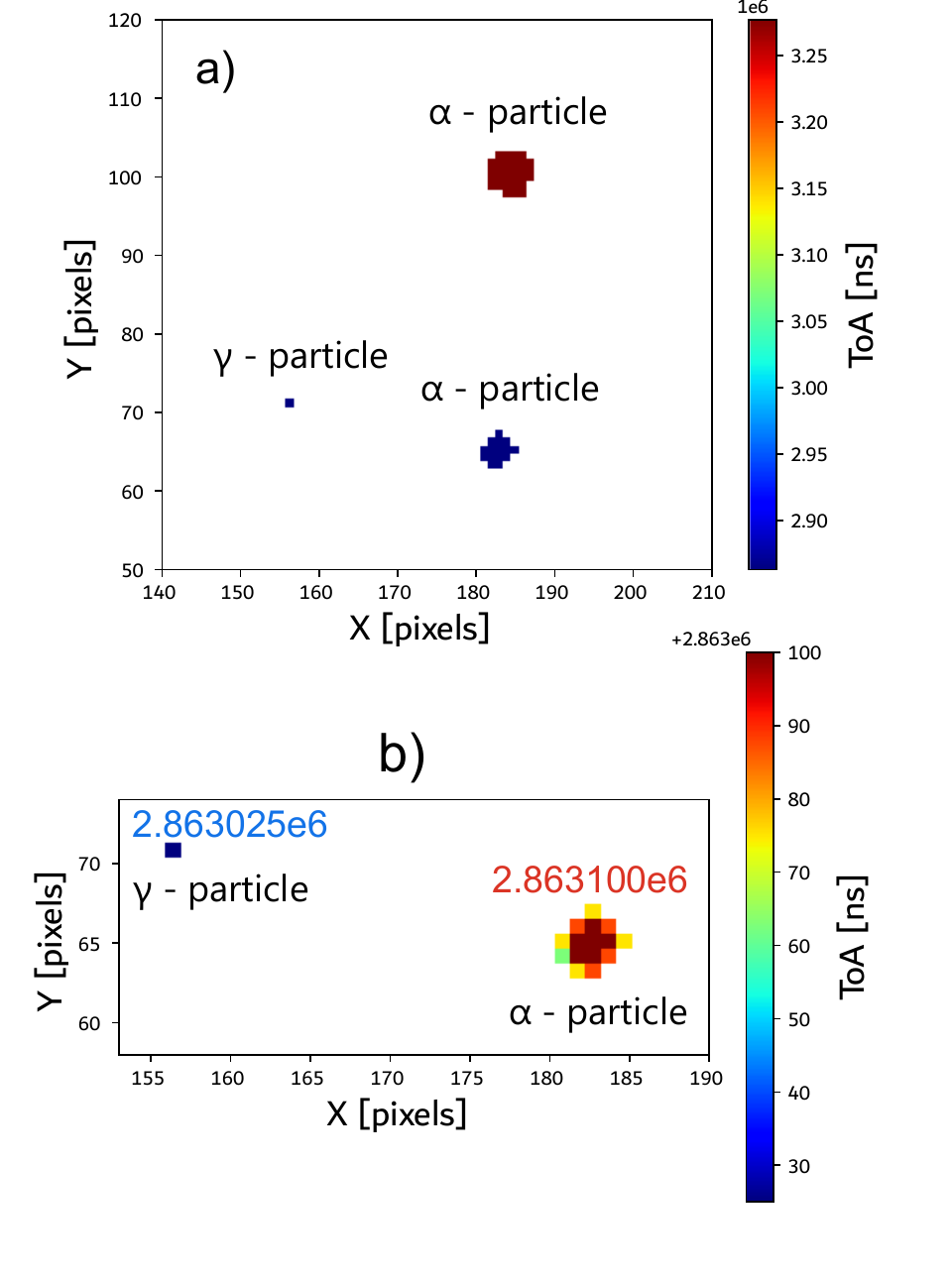}
	\caption{Typical frames acquired with the Timepix2 Lite detector with experimental. Figure (a) shows two $\alpha$ particles and $\gamma$ particle from $^{241}$Am. Figure (b) presents a zoomed-in view highlighting $\alpha$-$\gamma$ coincidence tracks and their Time-of-Arrival (ToA). }
	\label{fig:combined_frames}
\end{figure}

\begin{figure}[t]
	\centering
	\includegraphics[width=3.5in]{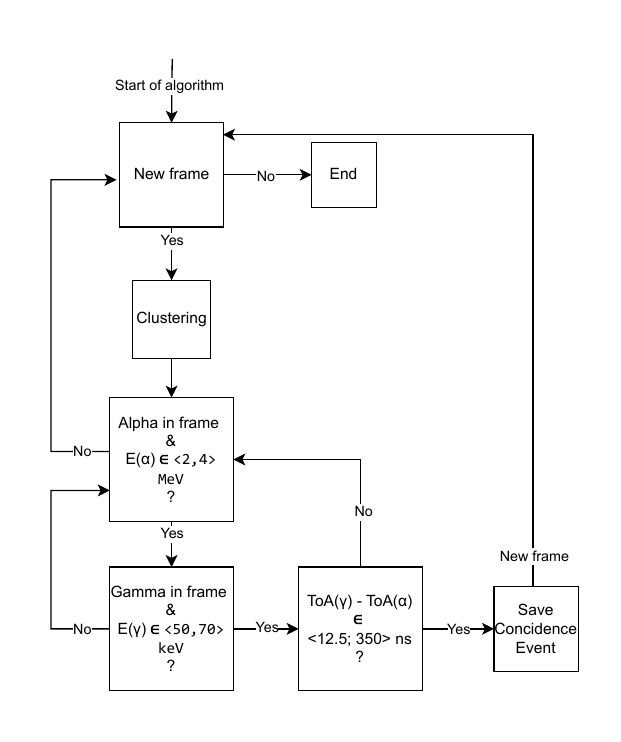}
	\caption{Simplified block diagram of the algorithm for alpha–gamma coincidence analysis in the measurement of the half-life of the second excited state in $^{237}$Np in air.}
	\label{Coincidence_anal}
\end{figure}

A typical frame acquired with the setup is presented in Fig.~\ref{fig:combined_frames} part (a). The figure displays particle tracks within the frame along with their reconstructed Time-of-Arrival in nanoseconds. A zoomed-in view of a  $\alpha$-$\gamma$ coincidence from this frame is shown in Fig. \ref{fig:combined_frames} part (b). Based on the particle classification scheme described in \cite{HOLY2008287}, $\alpha$-particle tracks can be distinguished from tracks produced by 59.5~keV $\gamma$-photons. This classification can be achieved, for example, by analyzing the track shapes and their morphological features. The simultaneously available information in each pixel about ToT and ToA from the Timepix2 Lite detector is key for designing an algorithm to correctly identify coincident $\alpha$-$\gamma$ events. A simplified block diagram of the proposed algorithm is shown in Fig.~\ref{Coincidence_anal}.


\par
\begin{figure}[t]
	\centering
	\begin{subfigure}[t]{0.48\textwidth}
		\centering
		\includegraphics[width=\linewidth]{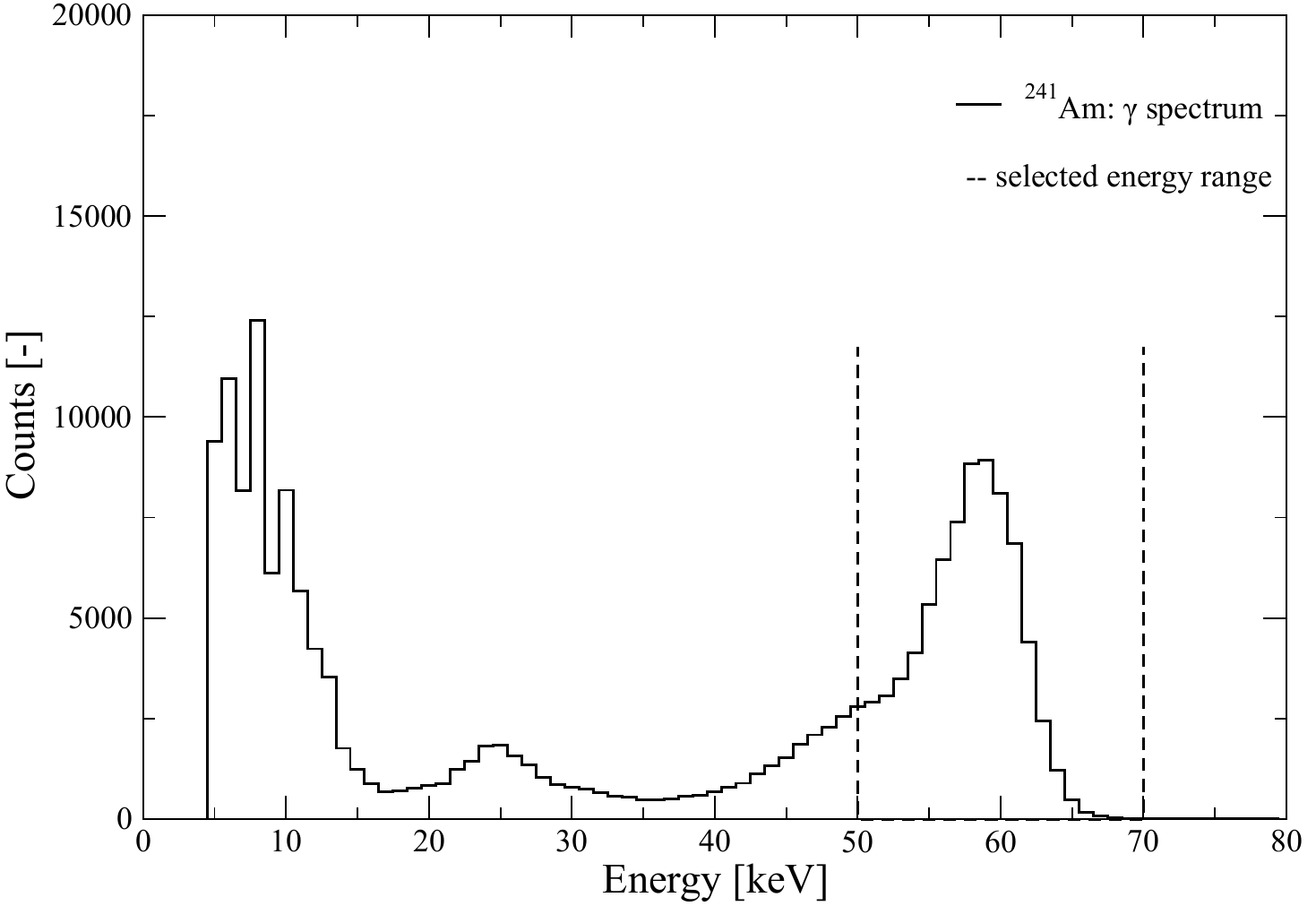}
		\caption{}
		\label{Gamma_spectrum}
	\end{subfigure}
	\hfill
	\begin{subfigure}[t]{0.48\textwidth}
		\centering
		\includegraphics[width=\linewidth]{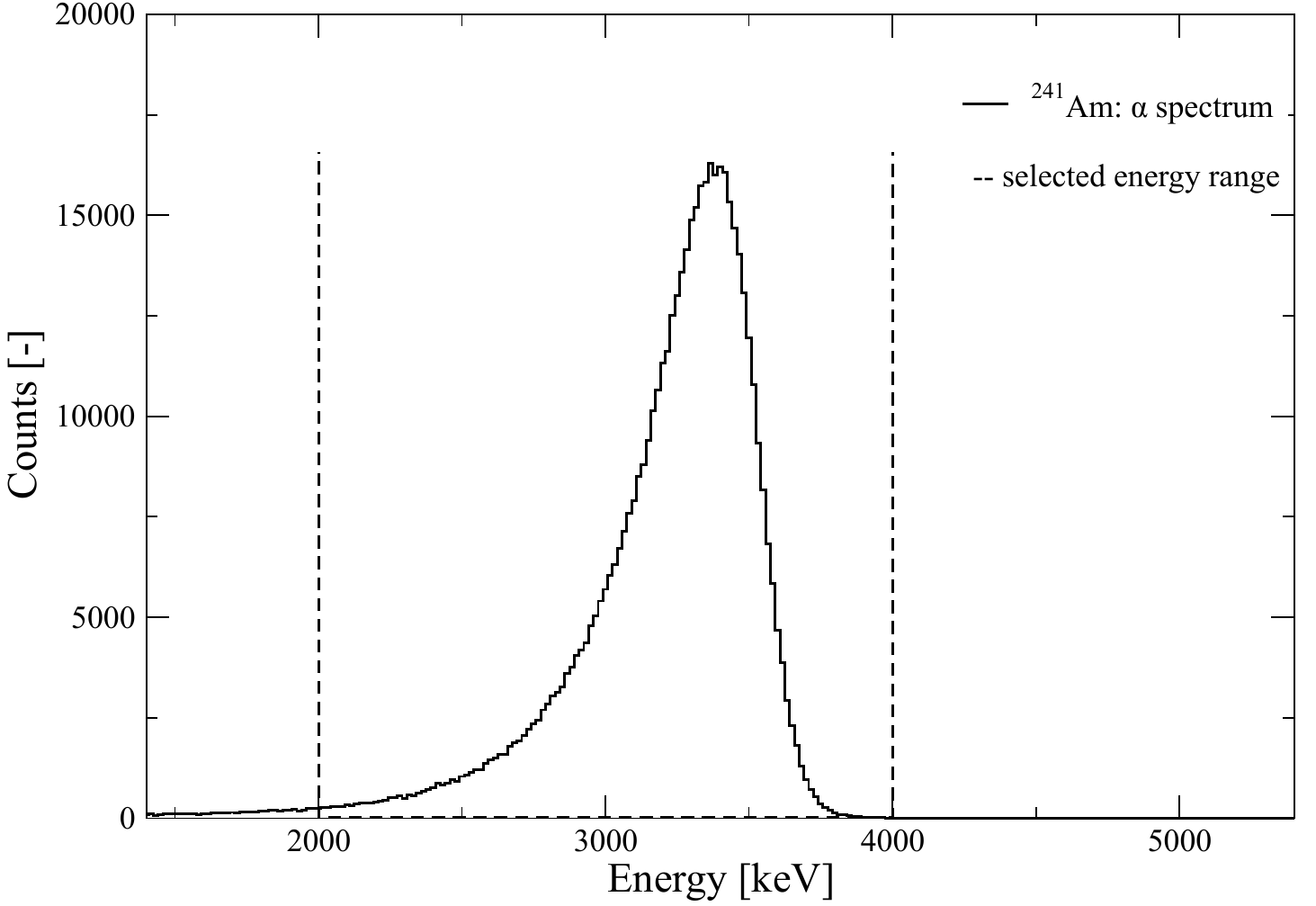}
		\caption{}
		\label{Alpha_spectrum}
	\end{subfigure}
	\caption{(a) $\gamma$-spectra and (b) $\alpha$-spectra of $^{241}$Am in air measured with Timepix2 Lite. The dashed lines indicate the selected energies used in the $\alpha$-$\gamma$ coincidence algorithm.}
	\label{Alpha_Gamma_spectrum}
\end{figure}

 The recorded frames are processed sequentially by the algorithm, frame by frame. For the purpose of this experiment, we restrict our algorithm to the analysis of individual frames. Furthermore, we do not consider cases in which one of the coincident events occurs at the end of one frame while the other appears at the beginning of the subsequent frame. According to the decay scheme of $^{241}$Am, the coincidence analysis is initiated by the detection of an $\alpha$-particle (transition A). When an $\alpha$-particle is detected within a frame and its energy falls within a predefined interval, ($E_{\alpha} \in [2,4]$ MeV), the event is flagged for further analysis. The energy loss (stopping power) of a 5.5~MeV $\alpha$-particle from $^{241}$Am traveling through air is approximately 1140~keV/cm. The sensor layer consists of a 0.5~$\mu$m thick aluminum backside contact and a 0.8~$\mu$m thick doped silicon dead layer, which together cause the $\alpha$-particle to lose a approximately 160~keV of its energy \cite{Mihai_2025}. Furthermore, energy losses of $\alpha$-particle due to the 1.6~$\mu$m thin gold foil on the surface of the radioactive source are also considered. This explains why the measured $\alpha$-spectrum in Fig.~\ref{Alpha_spectrum} exhibits a peak at lower energy compared to the nominal decay energy of $^{241}$Am.

If such an $\alpha$-particle is detected, the same frame is subsequently examined for the presence of a characteristic $\gamma$-photon with an energy of 59.5~keV (transition A, see Fig.~\ref{Am241_decay}). Specifically, $\gamma$-photon falling within the energy interval $E_{\gamma} \in [50,70]$~keV is considered as candidates for $\alpha$-$\gamma$ coincidence events. 
\par
\begin{figure}[t]
	\centering
	\includegraphics[width=3.5in]{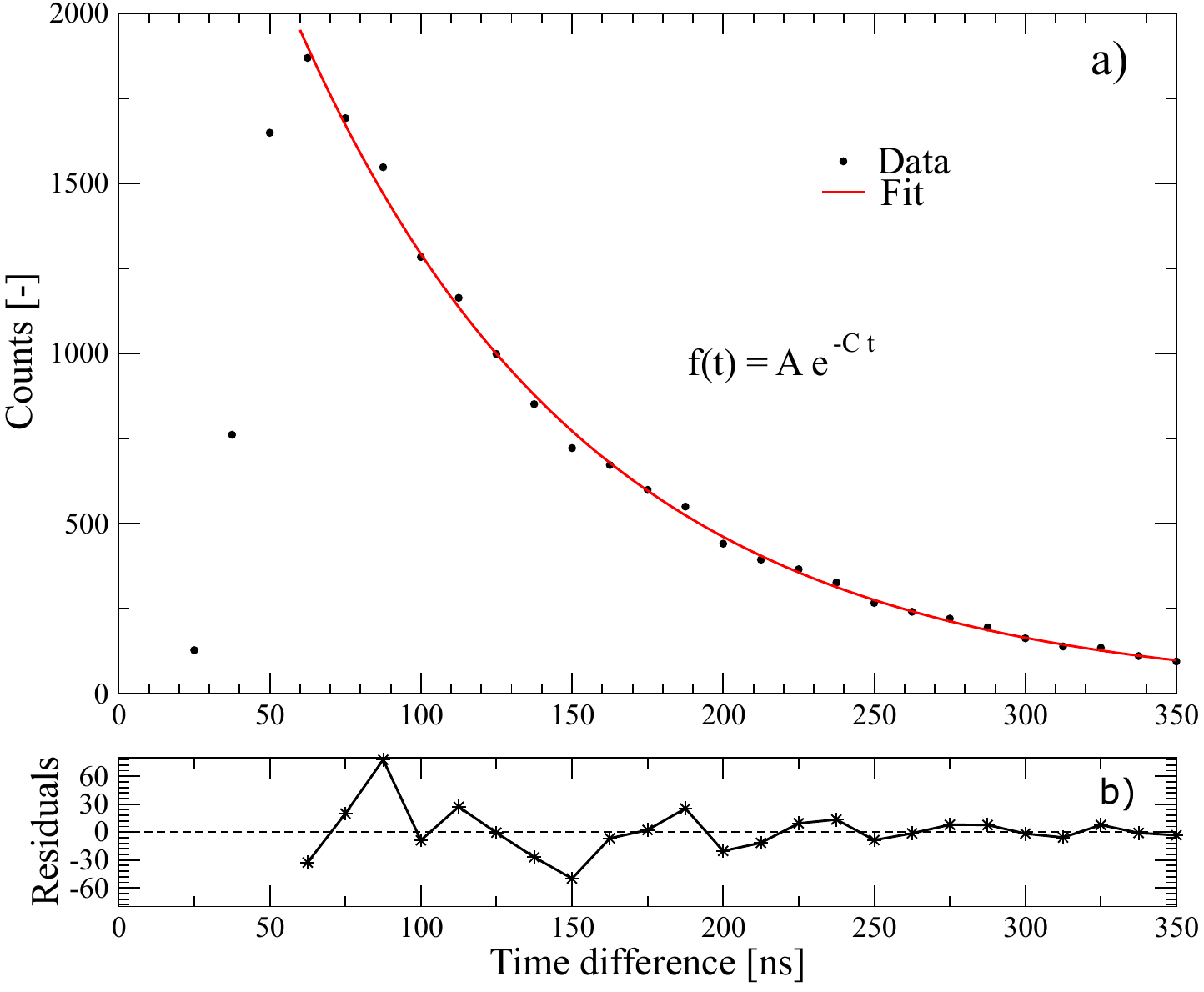}
	\caption{Decay curves of the 59.5~keV level measured with the Timepix2 Lite. An exponential fit (a) was applied to determine the half-lives. The residuals (b), representing the deviation of the measured data from the fitted curve.}
	\label{Half_life_result}
\end{figure}

 If such a pair of events is identified within a frame, and the difference in their Time-of-Arrival (ToA($\gamma$) - ToA($\alpha$)) value is in interval $[12.5;350]$~ns, the event is classified as an $\alpha$-$\gamma$ coincidence. The lower limit of the coincidence window was set to exclude events affected by the carrier drift time  in the 500~$\mu$m silicon layer and time-of-flight of $\alpha$-particle in the air. The upper limit of the coincidence window was set according to the previously reported half-life values from the literature (see Table \ref{tab:Neptunium}), chosen as 4–5 times the mean lifetime. These events are then collected in the $\alpha$-$\gamma$ time-difference spectrum, which is fitted with an exponential to determine the half-life as shown in Fig.~\ref{Half_life_result}.

\par
The corresponding energy spectra of $\alpha$ and $\gamma$ radiation, along with the selected energy ranges, are shown in Fig.~\ref{Alpha_Gamma_spectrum}.

\subsection{Results}
The time-difference spectra were fitted with an exponential function to determine the half-life t$_{1/2}$,
\begin{equation}
f(t) = A e^{-Ct},
\label{expo}
\end{equation}
where
\begin{equation}
C = \frac{\ln(2)}{t_{1/2}}
\end{equation}
Here, $A$ is the amplitude, proportional to the isotope activity. The influence of different fit range choices on the half-life determination is considered as systematic uncertainty. For the systematic error assessment, bounds of the fit range are varied. The intervals $I_k$ in which the fit was performed were chosen as: 
\[
I_k = [\,60 + 12.5k,\,350\,]~\text{ns}, \quad k = 0,1,2,3,4.
\]
From the employed fitting algorithm the standard deviation of the half-life fit for selected fit interval $I_k$, can be determined as:
\begin{equation}
	\sigma_{_{I_k}} = \frac{\partial t_{(1/2)_{I_k}}}{\partial C_{I_k}} = \frac{ln(2)}{C^2_{I_k}}\cdot SE(C_{I_k}) 
\end{equation}
where  $SE(C_{I_k})$ is standard error of the fitted parameter $C_{I_k}$.
The statistical error ($\delta_{stat.}$) is defined as the average of the errors from the individual fits. The systematic error ($\delta_{\mathrm{syst.}}$) is defined as the standard deviation of the half-life values obtained from the selected fitting intervals. The final error is defined as $\sqrt{\delta_{\mathrm{stat.}}^2 + \delta_{\mathrm{syst.}}^2}$.

The final measured half-life of the second nuclear excited state (state of 59.5 keV) in $^{237}$Np was determined to be:
\[
t_{1/2}^{\text{59.5 keV}} = (67.46 \pm 0.65_{\mathrm{stat.}} \pm 0.15_{\mathrm{syst.}})~\text{ns}
\]


\begin{table}[t]
	\centering
	\renewcommand{\arraystretch}{1.3} 
	\caption{\small\textsc{Half-life times of second excited state in $^{237}$Np found in literature.}}
	\begin{tabular}{lcc}
		\hline
		\textbf{Reference} & \boldmath$t_{1/2}$ (ns)  				\\
		\hline
		Nuclear Data Sheets (2006) 	\cite{BASUNIA2006} & 68.1(2)  	\\
		Vretenar et al. (2019) 	\cite{Vretenar_2019} & 67.7(1)  	\\
		Dutsov et al. (2021) 	\cite{DUTSOV2021} & 67.60(25)  	\\
		Santos et al. (2023) 	\cite{SANTOS2023} & 67.60(20)  	\\
		This work (2025) & 67.5	(7)							\\
		\hline
	\end{tabular}
	\label{tab:Neptunium}
\end{table}

\section{Future Measurements and Improvements}
\label{sec:Discussion}
The results are compared with previously reported half-lives of the 59.5~keV level, as summarized in Table~\ref{tab:Neptunium}. When comparing our educational measurement with previously reported results from literature, it can be observed that the measured value approaches those obtained in the most accurate contemporary measurements.
\par The measurement was performed in air, with the Timepix2 positioned approximately 1.5 cm from the $^{241}$Am source. The time-of-flight of the $\alpha$ particles in air should be reflected for future precise analysis. As illustrated on the left side of Fig.~\ref{Alpha_spectrum}, the rise time includes contributions from the drift time in silicon sensor layer and the time resolution of the Timepix2 detector.

\par The measurements were performed using a clock with a frequency of 80~MHz, corresponding to a clock period of 12.5~ns. If a detector with superior timing resolution, such as Timepix3 (Time-of-Arrival (ToA) clock of Timepix3 can be operated with a period of 1.56~ns \cite{Poikela_2014}), were employed, the results could approach the precision of currently reported half-life measurements, as reported in \cite{BergmannJelinek2022_PoHalfLives} and \cite{Bergmann2025Fe57HalfLives}.

\par An additional possible measurement of the half-life of the 59.5~keV level with this experimental setup is to use a X-ray fluorescence photon with an energy of 26.3~keV as a trigger for coincidence analysis instead of the $\alpha$-particle.

\section{Conclusion}
Timepix2 Lite introduces highly miniaturized readout interface for particle physics experiments, combining compact design with the advanced capabilities of the Timepix2 detector. Its integration with TrackLab software enables real-time interactive particle tracking, particle visualization, and sophisticated data processing, making it a powerful tool. The simultaneous measurement of Time-of-Arrival (ToA) and Time-over-Threshold (ToT), allows the implementation of innovative experiments, including demonstrated task of measuring the nanosecond-scale half-life of the second nuclear excited state in $^{237}$Np. These features highlight Timepix2 Lite’s contribution to both laboratory-based research and field-deployable nuclear instrumentation.
\appendices

\section*{Acknowledgment}

We thank the Medipix2 collaboration for developing and providing the Timepix2 ASIC. We
are grateful for the support of the PS/SPS coordinators, operators, and technicians. The authors would like to also thank Radu Mihai and Jindra Jelínek for their consultations and valuable comments.


\bibliographystyle{IEEEtran}
\bibliography{sample}  

\end{document}